\documentclass[a4paper]{article} \usepackage[utf8]{inputenc}
\usepackage[english]{babel} \usepackage[T1]{fontenc}
\usepackage{amsthm,amsmath,amssymb,amsfonts,stmaryrd,dsfont, mathtools}
\usepackage{mathpartir, etoolbox,xstring,ifthen, rotating}
\usepackage[hidelinks]{hyperref} \usepackage{tikz,tikz-cd,mytikz}
\usepackage{soul}
\usepackage{setspace,enumitem}
\usepackage{macros}
\usepackage{geometry}
\usepackage{multicol}
\usepackage{authblk}

\usetikzlibrary{decorations.pathmorphing}

\title{Monoidal weak $\omega$-categories\\as models of a type theory}

\author[1]{Thibaut Benjamin} \date{}
\affil[1]{Ecole Polytechnique, France, \href{mailto:thibaut.benjamin@polytechnique.edu}{thibaut.benjamin@polytechnique.edu}}

\makeatletter \newcommand{\neutralize}[1]{\expandafter\let\csname
  c@#1\endcsname\count@} \makeatother

\newtheorem{thm}{Theorem} \newtheorem*{thm*}{Theorem}
\newtheorem{prop}[thm]{Proposition} \newtheorem*{prop*}{Proposition}
\newtheorem{lemma}[thm]{Lemma} \newtheorem*{lemma*}{Lemma}
 \newtheorem{coro}[thm]{Corollary}

\theoremstyle{definition} \newtheorem{defi}[thm]{Definition}

\newtheorem{lemma-defi}[thm]{Definition/Lemma}
\newenvironment{lemma-defibis}[1]
{\renewcommand{\thethm}{\ref*{#1}$'$}%
  \addtocounter{thm}{-1}%
  \begin{lemma-defi}} {\end{lemma-defi}}

\theoremstyle{remark} \newtheorem{rem}[thm]{Remark}

\setenumerate[1]{label={(\arabic*)}} 
\begin{document}
\maketitle

\begin{abstract}
  Weak $\omega$-categories are notoriously difficult to define because of the very
intricate nature of their axioms. Various approaches have been explored, based
on different shapes given to the cells. Interestingly, homotopy type theory
encompasses a definition of weak $\omega$-groupoid in a globular setting, since
every type carries such a structure. Starting from this remark, Brunerie could
extract this definition of globular weak $\omega$\nobreakdash-groupoids,
formulated as a type theory. By refining its rules, Finster and Mimram have then
defined a type theory called $\CaTT$, whose models are weak
$\omega$-categories. Here, we generalize this approach to \emph{monoidal} weak
$\omega$-categories. Based on the principle that they should be equivalent to
weak $\omega$-categories with only one $0$-cell, we are able to derive a type
theory $\MCaTT$ whose models are monoidal categories. This requires changing the rules of the theory in order to
encode the information carried by the unique $0$-cell. The correctness of the
resulting type theory is shown by defining a pair of translations between our
type theory $\MCaTT$ and the type theory $\CaTT$. Our main contribution is to show that
these translations relate the models of our type theory to the models of the
type theory $\CaTT$ consisting of $\omega$-categories with only one $0$-cell, by
analyzing in details how the notion of models interact with the structural rules of both type
theories.

\end{abstract}

\setcounter{tocdepth}{2}
\tableofcontents

\newpage


Weak $\omega$-categories are algebraic structures occurring
naturally in modern algebraic topology and type theory. They consist of collections of
cells in every dimension, which can be composed in various ways. The
main difficulty in properly establishing a definition for those is due to the fact that the usual
coherence axioms imposed on composition, such as associativity, are
here relaxed and only supposed up to \emph{invertible} higher cells,
that we call \emph{witnesses} for these axioms. Moreover, these witnesses
themselves admit compositions, which satisfy axioms up to new
witnesses, and so on, making the compositions and their coherence axioms intricate. There have
been different approaches to propose a definition of weak
$\omega$-categories, which are summed up in a couple of
surveys~\cite{leinster2002survey,cheng2004higher}. These approaches
are based on various shapes, such as simplicial sets or opetopic
sets. In this article, we are interested in approaches based on
globular sets. Examples of such approaches can be found in the work of
Batanin~\cite{batanin1998monoidal} and
Leinster~\cite{leinster2004higher}, relying on the structure of
globular operad. Independently Maltsioniotis proposed an alternative
approach~\cite{maltsiniotis}, inspired by a definition of weak
$\omega$-groupoid (\ie~weak $\omega$-categories whose all cells are
invertible) proposed by Grothendieck~\cite{grothendieck1983pursuing},
and which is based on presheaves preserving some structure on a
well-chosen category. The two approaches have been proved equivalent
by Ara~\cite{ara2010groupoides}.

\paragraph{Type theoretical approach.}
An important observation which came along with the development of
homotopy type theory~\cite{hottbook} is the fact that the types in
Martin-Löf type theory, and in homotopy type theory carry a structure
of weak $\omega$-groupoid, where the higher cells are given by
identity
types~\cite{lumsdaine2009weak,van2011types,altenkirch2012syntactical}. This
allowed Brunerie to extract a minimal set of rules from homotopy type
theory for generating the weak
$\omega$-categories~\cite{brunerie2016homotopy}, that he could prove
to be equivalent to the definition of Grothendieck. Recently, Finster and
Mimram proposed a generalization of Brunerie's type theory to weak
$\omega$-categories~\cite{catt}, parallel to the generalization
Maltsiniotis proposed from Grothendieck definition. The type theory
they introduced is called $\CaTT$, and has been proved to be
equivalent to the definition of Maltsiniotis~\cite{benjamin2021globular}.

\paragraph{Monoidal categories.}
In this article, we are interested in monoidal weak
$\omega$-categories. These are categories equipped with a tensor
product allowing new ways to compose cells of every dimension. In
particular, one cannot compose the cells of dimension $0$ in weak
$\omega$-categories, but one can compose them in monoidal weak
$\omega$-categories. Moreover these new compositions are required to
satisfy axioms like associativity, but since it happens in a weak
setup, these axioms are again relaxed versions with witnesses in
higher dimensions.
This can be seen as a categorification of the notion of monoid.
Our goal here is to provide a variant of the \CaTT{} type theory in
order to describe monidal weak $\omega$-categories.

As noticed by Baez and Dolan \cite{baez1995higher,baezlecture},
monoidal weak $\omega$-categories should be equivalent to weak
$\omega$-categories with only one $0$-cell. In this correspondence,
there is a shift in dimension: the $0$-cells of the monoidal
$\omega$-category are the $1$-cells of the $\omega$-category with one
$0$-cell, and so on. The monoidal tensor product becomes the composition of
arrows of the $\omega$-category, and all its coherences are exactly
those satisfied by these arrows in the $\omega$-category.

Taking this correspondence as a starting point, we work out an
explicit type theory $\MCaTT$ whose models are monoidal categories, by
describing $\CaTT$ with the extra restriction that there should be only one
$0$-cells. To achieve this, we rely on the type theory $\CaTT$ and use its constituent to express the rules of the theory $\MCaTT$
We then define a pair of
translations back and forth between $\CaTT$ and $\MCaTT$, and use the
interaction of these translations with the structure of the type
theory to show that our proposed definition satisfies the
correspondence we started with.



\paragraph{Plan of the paper.}
In section 1, we introduce the type theory $\CaTT$ along with the
general tools we use to study type theories and their models. Then in section 2 we define
the type theory $\MCaTT$ along with a pair of translations between the theories $\CaTT$ and $\MCaTT$. We show that these translation exhibit a coreflective adjunction between the syntactic categories associated to the theory that lifts as an equivalence between a localization of the models of $\CaTT$ and the models of $\MCaTT$.


The author would like to thank Samuel Mimram and Eric Finster for their valuable discussions regarding the work presented in this article, as well as the reviewers for their helpful comments.
\section{Type theory for weak $\omega$-categories}
This section is dedicated to the type theory~$\CaTT$ introduced by
Finster and Mimram~\cite{catt} to describe weak $\omega$-categories.
The account we provide is introductory, and we refer the reader
to~\cite{benjamin2021globular} for a more in-depth presentation.
We first define our setting for type theory, and introduce a type
theory that describes globular sets. Then we extend this theory
with extra term constructors, to obtain the theory $\CaTT$.

\subsection{Type theories: notations and categorical semantics}
We establish the terminology and conventions that are used throughout
this article to manipulate type theories. Note that our method
consists in formulating a type theory to describe a structure, as
opposed to a developing said structure internally to Martin-Löf type
theory.

\paragraph{Expressions.}
A type theory manipulates various kind of objects, that we present
here along with the convention we use for naming them.
\begin{itemize}
\item \emph{variables} are elements of a given infinite countable set
  of variables, we denote them $x,y,z,\ldots$,
\item \emph{terms},are built out of variables and constructors, we
  denote them $t,u,\ldots$,
\item \emph{types}, which are built out of terms and constructors, we
  denote them $A,B,\ldots$,
\item \emph{contexts} are supported by association lists of the form
  $x:A$, they are denoted $\Gamma,\Delta,\ldots$,
\item \emph{substitutions} are supported by lists of mappings of the
  form $x\mapsto t$ where $x$ is a variable and $t$ a term, we denote
  them $\Gg,\Gd,\ldots$.
\end{itemize}
The constructors for terms and type depend on the theory we consider,
and we introduce them individually.  These all come with associated
set of variables, $\operatorname{Var}$ and that is the set of
variables needed to build it out.  We denote $\Var{t:A}$ for the union
of $\Var t$ and $\Var A$.  For contexts and substitutions, we always
have the following formulas
\begin{align*}
  \Var{\emptycontext} &= \emptyset
  & \Var{\GG,x:A} &= \Var \GG \cup \set{x}
  & \Var{\sub{}} &= \emptyset
  & \Var{\sub{\Gg,x\mapsto t}} &= \Var{\Gg}\cup \Var{t}
\end{align*}

\paragraph{Action of substitutions.}
We define the action of a substitution $\Gg$ on a type $A$ (\resp on a
term $t$) denoted $A[\Gg]$ (\resp $t[\Gg]$). These are defined
separately on each theory, but share the variable case in common:
\[
  x[\sub{}] = x \qquad\qquad x[\sub{\Gg,y\mapsto t}] = \left\{
    \begin{array}{ll}
      t & \text{If $x = y$} \\
      x[\Gg] & \text{Otherwise}
    \end{array}
  \right.
\]
This action provides a composition of substitutions given inductively
by
\begin{align*}
  \sub{}\circ\Gd &= \sub{}
  &
    \sub{\Gg,x\mapsto t}\circ\Gd
  &=
    \sub{\Gg\circ\Gd,x\mapsto t[\Gd]}
\end{align*}

\paragraph{Judgments.}
There are four kinds of judgments that are commons to all of our type
theories, they express the well-definedness of the previously
introduced objects:
\[
  \begin{array}{l@{\quad \rightsquigarrow
        \quad}l@{\qquad\quad}l@{\quad \rightsquigarrow \quad}l}
    \Gamma\vdash & \text{$\Gamma$ is a valid context} &
                                                        \Gamma\vdash t:A & \text{$t$ is a valid term of type $A$ in $\Gamma$}\\
    \Gamma\vdash A & \text{$A$ is a valid type in $\Gamma$} &
                                                              \GD\vdash \Gg:\GG & \text{$\Gg$ is a valid substitution from $\GD$ to $\GG$}
  \end{array}
\]
In order to distinguish, we sometimes refer to the raw syntax, or to
expressions for a syntactic entity which is not assumed to satisfy any
of the above judgment.

\paragraph{Basic rules.}
The following rules are common to all the type theories that we
consider
{\small
  \[
    \begin{array}{c@{\qquad}c}
      {\small \inferrule{\null}{\emptycontext\vdash}{\regle{ec}}}
    & {\small \inferrule{\Gamma\vdash A}{\Gamma,x:A\vdash}{\regle{ce}}} \\
      {\small \inferrule{\Gamma\vdash}{\Gamma\vdash\sub{}:\emptycontext}{\regle{es}}}
    & {\small \inferrule{\GD\vdash \Gg:\GG \\ \GG\vdash A \\ \GD\vdash
    t:A[\Gg]}{\GD\vdash\sub{\Gg,x\mapsto t}:\GG,x:A}{\regle{se}}}\\
      \multicolumn{2}{c}{
      {\small \inferrule{\Gamma\vdash \\ (x:A)\in\GG}{\Gamma\vdash x:A}{\regle{var}}}}
    \end{array}
  \]
}
Where in the rules \regle{ce} and \regle{se} we assume that
$x\notin\Var\GG$.

\paragraph{Syntactic category.}
We admit that for all the theories that we consider, these the action
of substitution makes the following rules admissible {\small
  \[
    \begin{array}{c@{\qquad}c@{\qquad}c}
      \inferrule{\GD\vdash \Gg: \GG \\ \GG\vdash A}{\GD\vdash A[\Gg]}
      &
        \inferrule{\GD\vdash \Gg: \GG \\ \GD\vdash t:A}{\GG\vdash t[\Gg] : A[\Gg]}
      &
        \inferrule{\GD\vdash \Gg: \GG \\ \GL\vdash \Gd : \GD}{\GL\vdash \Gg\circ\Gd : \GG}
    \end{array}
  \]} and that the action of substitutions is compatible with the
composition, thus making the composition associative
\begin{align*}
  A[\Gg][\Gd] &= A [\Gg\circ\Gd] & t[\Gg][\Gd] &= t[\Gg\circ\Gd]
  & \Gg\circ(\Gd\circ\delta) &= (\Gg\circ\Gd)\circ\delta
\end{align*}
Moreover, we can define an identity substitution, defined on a context
$\GG = (x_i:A_i)_{i\in I}$ by
$\id\GG = \sub{x_i\mapsto x_i}_{i\in I}$. One can show that whenever
$\GG\vdash$ is derivable, then so is $\GG\vdash \id\GG: \GG$ and that
whenever $\GD\vdash\Gg:\GG$, $\GG\vdash A$ and $\GG\vdash t:A$ hold,
the following equalities are satisfied.
\begin{align*}
  A[\id{\GG}] &= A & t [\id \GG] &= t & \Gg \circ \id\GG &= \Gg & \id \GD \circ \Gg &= \Gg
\end{align*}
Hence one can construct a \emph{syntactic category}
$\Syn{\mathfrak{T}}$ associated to a type theory $\mathfrak{T}$, whose
objects are the contexts of the type theory $\mathfrak{T}$, and whose
morphisms are the substitutions.

\paragraph{Properties.}
We admit that all our type theories also satisfy the following
properties, that one can prove by induction on the derivation
trees. We refer the reader to a formalization in
Agda\footnote{\url{https://github.com/thibautbenjamin/catt-formalization}}
where we define prove these properties for some of the theories.
\begin{prop}\label{prop:properties-tt}
  In all the theories we consider, the following rules are admissible
  {\small \[
      \begin{array}{c}
        \begin{array}{c@{\qquad\qquad}c@{\qquad\qquad}c@{\qquad\qquad}c}
          \inferrule{\GG\vdash A}{\GG\vdash}
          & \inferrule{\GG\vdash t:A}{\GG\vdash A}
          & \inferrule{\GD\vdash \Gg : \GG}{\GG\vdash}
          & \inferrule{\GD\vdash \Gg : \GG}{\GD\vdash}
        \end{array}
        \\
        \begin{array}{c@{\qquad}c@{\qquad}c}
          \inferrule{\GG,x:A\vdash \\ \GG \vdash B}{\GG,x:A \vdash B}
          & \inferrule{\GG,x:A \vdash \\ \GG\vdash t:A}{\GG,x:A\vdash t:B}
          & \inferrule{\GD,x:A \vdash \\ \GD\vdash \Gg:\GG}{\GD,x:A\vdash \Gg:\GG}
        \end{array}
      \end{array}
    \]} Moreover, for every derivable judgment $\GG\vdash A$ (\resp
  $\GG\vdash t:A$), we have $\Var A \subset \Var \GG$ (\resp
  $\Var t \subset \Var \GG$). For every derivable judgment
  $\GD\vdash \Gg:\GG$, we have $\Var \Gg \subset\Var \GD$, and writing
  $\Gg = \sub{x_i\mapsto t_i}_{0\leq i\leq n}$ and
  $\GG= (y_i : A_i)_{0\leq i\leq m}$, we have $n = m$ and for all $i$,
  $x_i = y_i$.
\end{prop}

\paragraph{Category with families.}
We use the formalism of \emph{categories with families}, introduced by
Dybjer~\cite{dybjer} as our categorical axiomatization of dependent
type theories. Denote $\Fam$ the category of families, where objects
are families of sets $(A_i)_{i\in I}$, and morphisms
$f:(A_i)_{i\in I}\to (B_j)_{j\in J}$ are pairs consisting of a
function $f:I\to J$ and a family of functions
$(f_i:A_i\to B_{f(i)})_{i\in I}$.
Given a category~$C$ equipped with a functor $T:\op{C}\to\Fam$. Given
an object $\Gamma$ of $C$, its image will be denoted
\[
  T\Gamma = \pa{\Tm^\Gamma_A}_{A\in\Ty^\Gamma}
\]
\ie $\Ty^\Gamma$ is the index set and $\Tm^\Gamma_A$ is an element of
the family. By analogy with a type theory, for a morphism
$\gamma : \Delta \to \Gamma$ an element $A\in\Ty^\Gamma$ and an
element $t\in\Tm^\GG_A$, we write $A[\gamma] = T\gamma(A)$ the image
of $A$ in $\Ty^\Delta$, and $t[\gamma] = T_A\gamma(t)$ the image of
$t$ in $\Tm^\GD_{A[\Gg]}$. With those notations, the functoriality
of~$T$ can be written as, for all composable morphisms $\Gg$, $\Gd$ in
$C$,
\begin{align*}
  A[\Gg\circ\Gd]
  &=
    A[\Gg][\Gd]
  &
    A[\id{}]&=A
  &
    t[\Gg\circ\Gd]
  &=
    t[\Gg][\Gd]
  &
    t[\id{}]&=t
\end{align*}
A \emph{category with families} (or \emph{CwF}) consists of a category
$C$ equipped with a functor as above $T:\op C\to\Fam$, such that $C$
has a terminal object, denoted $\emptycontext$, and that there is a
\emph{context comprehension} operation: given a context $\Gamma$ and
type $A\in\Ty^\Gamma$, there is a context $\pa{\Gamma,A}$, together
with a projection morphism $\pi:\pa{\Gamma,A} \to \Gamma$ and a term
$p\in\Tm^{\pa{\Gamma,A}}_{A[\pi]}$, such that for every morphism
$\sigma : \Delta\to\Gamma$ in $C$ together with a term
$t \in\Tm^{\Delta}_{A[\sigma]}$, there exists a unique morphism
$\sub{\sigma,t} : \Delta\to\pa{\Gamma,A}$ such that
$p[\sub{\sigma,t}] = t$:
\[
  \begin{tikzcd}
    &(\Gamma,A)\ar[d,"\pi"]\\
    \Delta\ar[ur,dotted,"\sub{\sigma,t}"]\ar[r,"\sigma"']&\Gamma
  \end{tikzcd}
\]
We call an arrow of the form $\pi:(\GG,A)\to\GG$ a \emph{display
  map}. The following result is well-known and allow for giving
structure to the syntactic category of a type theory.
\begin{prop}
  The syntactic category of a type theory is endowed with a structure
  of category with families. For every context $\GG$, $\Ty^\GG$ is the
  set of derivable types in $\GG$ and $\Tm^\GG_A$ is the set of
  derivable terms of type $A$ in $\GG$.
\end{prop}


\paragraph{Models of a category with families.}
The structure of category with families enforces some compatibility
between context comprehension and the action of morphisms on the type.
\begin{lemma}\label{lemma:pullback-display-map}
  In a category with families~$C$, for every morphism
  $f:\Delta\to\Gamma$ in~$C$ and $A\in\Ty^\Gamma$, the following
  square is a pullback
  \[
    \begin{tikzcd}
      (\Delta,A[f])\ar[d,"\pi'"'] \ar[r,"\sub{f\circ\pi',p'}"]& (\Gamma,A)\ar[d,"\pi"]\\
      \Delta \ar[r,"f"'] & \Gamma
    \end{tikzcd}
  \]
  where $\pi':(\Delta,A[f])\to\Delta$ and
  $p'\in\Tm^{(\Delta,A[f])}_{A[f][\pi']}$ are obtained by context
  comprehension.
\end{lemma}
This is essentially a reformulation of the universal property of the
context comprehension, and was observed by Dybjer~\cite{dybjer}. The
entire formalism of categories with families can be seen as an
axiomatization of a category with a split choice of pullbacks along
the display maps.


\begin{defi}[Models of a category with families]
  Consider a category with families $C$, we say that a functor
  $F : C \to \Set$ is a \emph{(set-theoretic) model} of $C$ if it
  sends all the pullbacks along display maps in $C$ onto pullbacks in
  $\Set$.
\end{defi}

\paragraph{Morphisms of category with families.}
We call a \emph{morphism of categories with families} a functor that
preserves the terminal object and the context comprehension on the
nose. Since the pullbacks are computed from this structure, the
morphisms of categories with families preserve the pullbacks along
display maps.

\subsection{Globular sets}
Globular weak $\omega$-categories are supported by \emph{globular
  sets}, which are presheaves over the category of globes $\G$,
generated by $
\begin{tikzcd}
  0 \ar[r,shift left = 1, "s"] \ar[r,shift right = 1, "t"'] & 1
  \ar[r,shift left = 1, "s"] \ar[r,shift right = 1, "t"'] & 2
  \ar[r,shift left = 1, "s"] \ar[r,shift right = 1, "t"'] & \cdots
\end{tikzcd}
$ (with $ts = ss$ and $st = tt$). We denote $\GSet = \widehat\G$ the
category of globular sets, 
and in this category, we call the $n$-dimensional \emph{disk} the
representable object represented by $n$ and denote it $D^n$.


\paragraph{Type theory for globular sets.}
Following~\cite{catt} we first give a type theoretic description of
the theory $\Glob$ describing globular sets. This theory has two type
constructors $\Obj$ and $\Hom{}{}{}$ and no term
constructor. Intuitively, $\Obj$ is the types of the objects of the
set, and given a type $A$ and two terms $t$, $u$, the type $\Hom{A}tu$
is the type of higher cells from $t$ to $u$. Define the variables of a
type, as well as the action of substitutions as follows
\begin{align*}
  \Var\Obj &= \emptyset
  & \Var{\Hom Atu} &= \Var A \cup \Var t \cup \Var u
  & \Obj[\Gg] &= \Obj
  & (\Hom Atu)[\Gg] &= \Hom{A[\Gg]}{t[\Gg]}{u[\Gg]}
\end{align*}
These constructors are subject to the following introduction rules
{\small
  \[
  \inferrule{\Gamma\vdash}{\Gamma\vdash\Obj}{\regle{$\Obj$-intro}}
  \qquad\qquad\qquad \inferrule{\Gamma\vdash t:A \\ \Gamma\vdash
    u:A}{\Gamma\vdash \Hom Atu}{\regle{$\Hom{}{}{}$-intro}}
\]}
Since there are no term constructors, the only terms in the theory
$\Glob$ are variables. A self contained description of this theory is
provided in Appendix~\ref{sec:app-glob}.

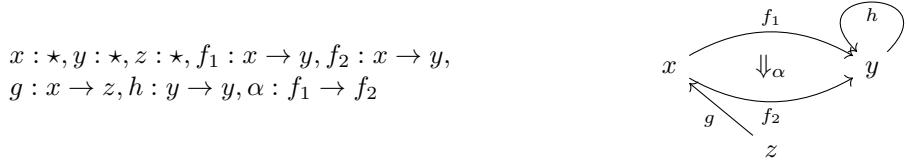
\begin{figure}
  \centering
  \begin{multicols}{2}
    \[
      \begin{array}{l}
        x:\Obj,y : \Obj,z : \Obj,  f_1 :\Hom{}xy, f_2 : \Hom{}xy,\\
        g : \Hom{}xz, h : \Hom{}yy, \alpha :
        \Hom{}{f_1}{f_2}
      \end{array}
    \]
    \[
      \begin{tikzcd}
        x \ar[rr, bend left = 30, "f_1"]\ar[rr, bend right = 30,
        "f_2"']\ar[rr, phantom, "\Downarrow_{\alpha}"] & & y\ar[loop, "h"] \\
        & z\ar[ul, "g"] &
      \end{tikzcd}
    \]
  \end{multicols}
  \caption{A context and its corresponding globular set}
  \label{fig:corresp-glob}
\end{figure}

\paragraph{Semantics of the theory $\Glob$.}
Here, we present results about the theory $\Glob$ without proofs; a
detailed presentation can be found
in~\cite{catt,benjamin2021globular}. The following result
characterizes the syntactic category of $\Glob$ ans is illustrated in Fig.~\ref{fig:corresp-glob}.
\begin{prop}[{\cite[Prop. 11]{catt},\cite[Th. 16]{benjamin2021globular}}]
  \label{prop:syn-glob}
  The syntactic category of the theory $\Glob$ is equivalent to the
  opposite of the category of finite globular sets.
\end{prop}

\begin{prop}[{\cite[Prop 13]{catt},\cite[Th. 22]{benjamin2021globular}}]
  The category of models of $\Glob$ is equivalent to the category of
  globular sets.
\end{prop}


\paragraph{Disk contexts.}
The category of finite globular sets contains in particular the
representable objects $\D^n$, we also denote $\D^n$, and call
\emph{disk contexts}, the corresponding contexts in $\Glob$ given by
Proposition~\ref{prop:syn-glob}.  In low dimensions, they are given
(up to renaming of their variables), by
\begin{align*}
  \D^0 &= (x : \Obj)&
                      \D^1 &=  (x:\Obj,y:\Obj,f:\Hom{} xy) &
                                                             \D^2 &= (x:\Obj,y:\Obj,f:\Hom{} xy, g:\Hom{} xy, \alpha : \Hom{}fg)
\end{align*}

\subsection{The type theory $\CaTT$}
\label{sec:catt}
In order to axiomatize the weak $\omega$-category structure, we add
new terms, that we call operations and coherences. This can be done by
adding term constructors to the theory $\Glob$. To express the rules
for these constructors, we introduce the \emph{ps-contexts}, which are
a particular class of ps-contexts.

\paragraph{Ps-contexts.} We define \emph{ps-contexts} to be a
particular class of contexts in the theory $\Glob$ that we describe
using rules of a type-theoretic flavor. We thus introduce two new
judgment kinds
\[
  \begin{array}{l@{\quad\rightsquigarrow\quad}l}
    \Gamma\vdashps & \text{the context $\Gamma$ is a ps-context}\\
    \Gamma\vdashps x:A & \text{$\Gamma$ is a partial ps-context with dangling variable $x$}
  \end{array}
\]
The second judgment is just a convenient auxiliary, where the dangling
variable indicates the unique variable on which one is allowed to glue
a cell at the next step.  These two judgments are subject to the
following rules {\small
  \[
    \begin{array}{c@{\quad}c@{\quad}c@{\quad}c}
      \inferrule{\null}{(x:\Obj)\vdashps x:\Obj}{\regle{pss}} &  \inferrule{\Gamma\vdashps f:\Hom Axy}{\Gamma\vdashps y:A}{\regle{psd}} &
                                                                                                                                          \inferrule{\Gamma\vdashps x:A}{\Gamma,y:A,f:\Hom Axy\vdashps f: \Hom Axy}{\regle{pse}} & \inferrule{\Gamma\vdashps x:\Obj}{\Gamma\vdashps}{\regle{ps}}
    \end{array}
  \]}

A ps-context $\Gamma$ comes equipped with notion of \emph{$i$-source}
$\src i \GG$ and \emph{$i$-target} $\tgt i \GG$ for every number
$i\in\N$, defined inductively on its structure:
\[
  \src i {(x:\Obj)} = (x:\Obj)
  \qquad\qquad\partial^-_i(\Gamma,y:A,f:\Hom{}xy) = \left\{
    \begin{array}{l@{\quad}l}
      \partial^-_i\Gamma & \text{if $\dim A \geq i$}\\
      \partial^-_i\Gamma,y:A,f:\Hom{}xy & \text{otherwise}
    \end{array}
  \right.
\]
\[
  \tgt i {(x:\Obj)} = (x:\Obj) \qquad\qquad
  \partial^+_i(\Gamma,y:A,f:\Hom{}xy) = \left\{
    \begin{array}{l@{\quad}l}
      \partial^+_i\Gamma & \text{if $\dim A > i$}\\
      \operatorname{drop}(\partial^+_i\Gamma),y:A & \text{if $\dim A = i$}\\
      \partial^+_i\Gamma,y:A,f:\Hom{}xy & \text{otherwise}
    \end{array}
  \right.
\]
\noindent where $\operatorname{drop}(\Gamma)$ is the context $\Gamma$
with its last variable removed, \ie
$\operatorname{drop}(\Gamma,x:A)=\Gamma$. We write
$\partial^{\pm} \GG = \partial^{\pm}_{\dim\GG - 1}$. With these
conventions, $\partial^-(x:\Obj)$ and $\partial^+(x:\Obj)$ are not
defined.
\begin{rem}
  Under the correspondence of Proposition~\ref{prop:syn-glob}, the
  ps-contexts correspond to a class of finite globular sets called
  \emph{pasting schemes}, which can be described in several different
  way and play a significant role in weak $\omega$-category
  theory~\cite{batanin1998monoidal,leinster2004higher,maltsiniotis}.
  Intuitively they are the globular sets that can be completely
  composed in an essentially unique way, and they are used to index
  all the operations. More details about their relation to ps-contexts
  can be found in~\cite[Th.73]{benjamin2021globular}.
\end{rem}

\paragraph{Syntax of the theory $\CaTT$.} The syntax is obtained by
adding two term constructors $\cohop$ and $\coh$, do the theory
$\Glob$. A term expression in this theory is thus defined to be either
a variable or of the form $\cohop_{\GD,A}[\Gd]$ or
$\coh_{\GD,A}[\Gd]$, where in both the latter cases $\GD$ is a
context, $A$ is a type and $\Gd$ is a substitution. To simplify the
notations, we denote $\constr$ to be either $\cohop$ or $\coh$.The set
of variables and the application of substitutions are defined by the
formulas
\begin{align*}
  \Var{\constr_{\GG,A}[\Gg]} &= \Var{\Gg}
  &\constr_{\GG,A}[\Gg][\Gd] &= \constr_{\GG,A}[\Gg\circ\Gd]
\end{align*}

\paragraph{Side conditions.}
The introduction rules for the term constructors $\cohop$ and $\coh$
have to verify some side conditions regarding the variables that are
used in the type that we derive. Intuitively the introduction rule for
the constructor $\cohop$ creates witnesses for \emph{operations}, for
instance the composition, or whiskering, which a priori have no reason
to be invertible, whereas the introduction rule for the constructor
$\coh$ creates witnesses for \emph{coherences}, as for instance the
associators, which are always weakly invertible. In order to simplify
the notations, we encapsulate the requirements for these rules along
with their side conditions in new judgments $\GG\vdashop A$ and
$\GG\vdasheq A$ that we interpret as stating that $A$ defines a valid
operation (\resp coherence) in the context $\GG$.  These two judgments
are subject to the following derivation rules, which express all the
requirements for the introduction rules of the constructors $\cohop$
and $\coh$
\[
  {\small \inferrule{\Gamma\vdashps \\ \src\Gamma\vdash t : A \\
    \tgt\Gamma\vdash u:A}{\Gamma\vdashop \Hom Atu}} \quad \left\{
    \begin{array}{l}
      \Var{t:A} = \Var{\src\Gamma}\\
      \Var{u:A} = \Var{\tgt\Gamma}
    \end{array}
  \right.
\]
\[
  {\small \inferrule{\Gamma\vdashps \\ \Gamma\vdash t:A \\ \Gamma\vdash
    u:A}{\Gamma\vdasheq A}} \quad \left\{
    \begin{array}{l}
      \Var{t:A} = \Var{\Gamma}\\
      \Var{u:A} = \Var{\Gamma}
    \end{array}
  \right.
\]
\noindent Note that whenever $\GG\vdashop A$ is derivable, then a
top-dimensional variable of $\GG$ cannot appear in $A$, whereas
whenever $\GG\vdasheq A$ is derivable, a top-dimensional variable of
$\GG$ has to appear in $A$, hence these two rules are mutually
exclusive.

\paragraph{Operations and coherences.}
We can now give the introduction rules for the new term constructors
$\coh$ and $\cohop$.
{\small
  \[
  \begin{array}{c@{\qquad\qquad\qquad}c}
    \inferrule{\Gamma\vdashop A
    \\\Delta\vdash\Gg:\Gamma}{\Delta\vdash \cohop_{\Gamma,A}[\Gg]:
    A[\Gg]}{\regle{$\cohop$-intro}}
    &
      \inferrule{\Gamma\vdasheq A \\ \Delta\vdash\Gg:\Gamma}
    {\Delta\vdash\coh_{\Gamma,A}[\Gg] : A[\Gg]}{\regle{$\coh$-intro}}
  \end{array}
\]}
The resulting type theory obtained by adding these rules is called
$\CaTT$. We present all the rules of the type theory $\CaTT$ together
in Appendix~\ref{sec:app-catt}.

\paragraph{Interpretation.} The ps-contexts intuitively represent all
the context in which there ought to be an essentially unique
definition (in a weak situation, they have a contractible set worth of
composition). The rule~\regle{\cohop-intro} enforces a composition to
exist for every ps-context, while the rule~\regle{\coh-intro} enforces
that any two composition of a ps-context are related, together, they
provide a directed formulation of the idea of contractibility. Note
that our formulation of the rule \regle{\coh-intro} is slightly
different than the one introduced by Finster and
Mimram~\cite{catt}. Actually the two can be proven equivalent, but the
proof is surprisingly involved~\cite[Coro. 103]{benjamin2020type}.

\paragraph{Examples.}
We give a few examples of derivations in this system illustrating how
it describes weak $\omega$-categories. This shows the introduction of
a new operation and a new coherence, and emphasizes the role of the
substitutions taken as their arguments.
\begin{itemize}
\item Composition: In $\CaTT$ one can use an operation to derive a
  witness for composition of $1$-cells. Start by considering the
  context
    $\Gamma_ {\texttt{comp}} =
    (x:\Obj,y:\Obj,f:\Hom{}xy,z:\Obj,g:\Hom{}yz)$.
  One can check that $\Gamma_{\texttt{comp}}\vdashps$, and compute its
  source $\src{\Gamma_{\texttt{comp}}} = (x:\Obj)$ and target
  $\tgt{\Gamma_{\texttt{comp}}} = (z:\Obj)$. Thus, the judgment
  $\Gamma_{\texttt{\comp}}\vdashop \Hom{}xz$ is derivable. Now
  considering with two terms $v,w$ in a context $\GG$ such that
  $\Gamma\vdash v:\Hom{}{u}{u'}$ and $\Gamma\vdash w:\Hom{}{u'}{u''}$
  (\ie{}two composable $1$-cells $u$ and $v$), we define the
  substitution
  $\Gg = \sub{x\mapsto u,y\mapsto u',f\mapsto v,z\mapsto u'',g\mapsto
    w}$. The judgment $\GG\vdash \Gg:\Gamma_{\texttt{\comp}}$ is then
  derivable, thus so it
  $\Gamma\vdash\cohop_{\Gamma_{\texttt{comp},\Hom{}xz}}[\Gg] :
  \Hom{}{u}{u''}$. We denote this term with the simpler and more usual
  notation $\Gamma\vdash \texttt{comp}\ v\ w : \Hom{}{u}{u''}$.
\item Associativity: Similarly, one can define the context
  \[
    \Gamma_ {\texttt{assoc}} =
    (x:\Obj,y:\Obj,f:\Hom{}xy,z:\Obj,g:\Hom{}yz,w:\Obj,h:\Hom{}zw)
  \]
  and check that $\Gamma_{\texttt{assoc}}\vdashps$, and
  $\Gamma_{\texttt{assoc}} \vdasheq \Hom{}{\texttt{comp}\
    (\texttt{comp}\ f\ g)\ h} {\texttt{comp}\ f\ (\texttt{comp}\ g\
    h)} $ are both derivable. Whenever a context $\Gamma$ defines
  three terms $u,v,w$ that are composable $1$-cells, the
  rule~\regle{$\coh$-intro} provides a witness for the associativity
  of their compositions, denoted
  \[
    \Gamma\vdash\texttt{assoc}\ u\ v\ w : \Hom{}{\texttt{comp}\
      (\texttt{comp}\ u\ v)\ w} {\texttt{comp}\ u\ (\texttt{comp}\ v\
      w)}
  \]
\end{itemize}

\paragraph{Models.}
For the purpose of this article, we define the category of weak
$\omega$-categories to be the category of models of $\CaTT$, and we
refer the reader to~\cite{benjamin2021globular} for a formal
exploration of how they relate to other known definitions of weak
$\omega$-categories.  We can however explain informally why this
definition is sensible: Considering a model $F:\Syn{\CaTT}\to\Set$, we
define its set of objects to be $F(D^0)$, and given two objects
$x,y\in F(D^0)$, the set of $1$-cells between $x$ and $y$ in $F$ to be
$\operatorname{Hom}_F(x,y) = \set{f\in F(D^1), F(s)(f)=x, F(t)(f)=y}$
where $D^1\vdash s:D^0$ is the source substitution and
$D^1\vdash t:D^0$ is the target substitution. Two $1$-cells
$f\in\operatorname{Hom}_F(x,y)$ and $g\in\operatorname{Hom}_F(y,z)$
define a unique element of $F(\GG_{\texttt{comp}})$. Then, the
constructor \texttt{comp} defines a substitution
$\GG_{\texttt{comp}}\to D^1$, which induces the element of $F(D^1)$
corresponding to the composition of $f$ and $g$. This observation can
be generalized, so that all derivable terms correspond to additional
algebraic structure on the globular set induced by $F$.


\paragraph{Syntactic category.} We have seen that the category
$\Syn{\Glob}$ consists in the finite globular sets, which are also the
finitely generated ones since the representable are finite. Similarly
the category $\Syn\CaTT$ can be conceived as the full subcategory of
$\Mod{\Syn{\CaTT}}$ whose objects are freely finitely generated.
However, stating this formally is not straightforward since it
requires adapting the notion of polygraph~\cite{burroni1993higher} (also
known as computads~\cite{street1976limits}). These results can
be compared with the Gabriel-Ulmer duality~\cite{gabriel2006lokal}.



\section{The type theory $\MCaTT$ for monoidal weak $\omega$-category}\label{sec:mcatt}
We now present a new type theory $\MCaTT$, which is an original contribution of
this article, and which aims at describing monoidal weak $\omega$-categories.
The main intuition behind this definition is to start from the idea that a monoidal 
weak $\omega$-category with a single object, and enforce this constraint on the
theory $\CaTT$. This induces a shift in dimension, so we refer to this new unique object
as being of ``dimension $-1$'', in such a way that the cells of dimension $0$ of the monoidal
categories really are the objects.


\paragraph{The subcategory of context with one object.}
A natural candidate for the syntactic category of the theory of
categories with one object is the full subcategory of $\Syn{\CaTT}$
whose objects are the contexts containing exactly one cell of type
$\star$. We denote this category $\Syn{\CaTT,\bullet}$. It is not
clear however that this category can be equipped with a structure of
category with families. We introduce the category $\MCaTT$, and prove
later that its syntactic category is equivalent to
$\Syn{\CaTT,\bullet}$. This answers the above question, by
constructing an actual theory presenting $\Syn{\CaTT,\bullet}$.


\subsection{Globular sets with a unit type}
We first define a variant of the type theory $\Glob$, that we call
$\Glob_{\unit}$, in order to handle the dimension shift that happens.

\paragraph{Unit type.}
The type theory $\Glob_{\unit}$ is obtained from the theory $\Glob$ by
formally replacing the type $\Obj$ by a unit type that we denote
$\unit$, together with a constant that we denote \texttt{()}, which
represent a term of this unit type. We keep the constructor
$\Hom{}{}{}$. These new constructors have the following set of
variables and action under substitutions
\begin{align*}
  \Var\unit &= \emptyset & \Var{\texttt{()}} &= \emptyset  & \unit[\Gg] &= \unit & \texttt{()}[\Gg] &= \texttt{()}
\end{align*}
The type introduction rules for the theory $\Glob_{\unit}$ are the
following {\small
  \[
    \inferrule{\GG\vdash}{\GG\vdash\unit}{\regle{$\unit$-intro}}
    \qquad\qquad \inferrule{\GG\vdash}{\GG\vdash
      \texttt{()}:\unit}{\regle{$\texttt{()}$-intro}} \qquad\qquad
    \inferrule{\GG\vdash A \\ \GG\vdash t:A \\ \GG\vdash
      u:A}{\GG\vdash \Hom Atu}{\regle{$\Hom{}{}{}$-intro}}
  \]} We also add a computation rule to this theory, that postulates a
new definitional equality. We use the symbol $\equiv$ to denote such
definitional equalities. We require the two following rules, so that
typing respects definitional equality and that $\unit$ is a unit type.
{\small
  \[
    \inferrule{\GG\vdash t:A \\ \GG\vdash A\equiv B}{\GG\vdash t:B}{} \qquad\qquad
    \inferrule{\GG\vdash x:\unit}{\GG\vdash
      x\equiv\texttt{()}:\unit}{\regle{$\eta_{\unit}$}}
  \]}
We refer the reader to Appendix~\ref{app:glob-unit} for a summary of all
the rules of the theory

\paragraph{Theories with definitional equalities}
The addition of definitional equality complicates the theory quite
substantially in general. A slight technical problem is that the rule
\regle{$\eta_{\unit}$} makes the set of variables of a term not
invariant. This is easily solved since we do not use this set in the
presence of the unit type. A more profound issue is that one needs to
account for this equality in the definition of the syntactic category.
This is why we now take the set of objects to be the contexts up to
definitional equality, and the morphisms to be substitutions up to
definitional equality. We also consider the set of terms and types
associated to a context to be up to definitional equality. The last
difficulty is that adding definitional equality may break the
decidability of type checking. We admit that in our case, we can
actually define a normal form by rewriting any term
$\GG\vdash t:\unit$ into the term \texttt{()}. This ensures that the
type checking is decidable. From now on, we always implicitly
normalize all the expression we consider and thus eliminate all the
difficulties due to definitional equality.

\paragraph{Notations for mixing theories.} When it is ambiguous, we
write $\vdash_{\mathcal{T}}$ for judgments that are derivable in the
type theory $\mathcal{T}$. In order to simplify the formulation, we
allow for rules that combine judgments coming from different theory,
in particular, we may say for example that any of the following rule
is admissible to express the fact that we can construct a derivation
of the lower judgment in the theory $\mathcal{T}'$ from a derivation
of the upper judgment in the theory $\mathcal{T}$.
{\small
  \[
    \inferrule{\GG\vdash_{\mathcal{T}}}{\GG'\vdash_{\mathcal{T}'}}
    \qquad\qquad
    \inferrule{\GG\vdash_{\mathcal{T}}A}{\GG'\vdash_{\mathcal{T}'}A'}
    \qquad\qquad
    \inferrule{\GG\vdash_{\mathcal{T}}t:A}{\GG'\vdash_{\mathcal{T}'}t':A'}
    \qquad\qquad
    \inferrule{\GD\vdash_{\mathcal{T}}\Gg:\GG}{\GD'\vdash_{\mathcal{T}'}
      \Gg':\GG'}
  \]}

\paragraph{Dimension and type of terms.}
We define the dimension of a type in the theory $\Glob_{\unit}$ by
induction: $\dim{\unit} = -2$ and $\dim{\Hom Atu} = \dim A + 1$.  The
dimension of a term $\GG\vdash t:A$ is $\dim t = 1 + \dim A$. Note
that for every context $\GG$, there is only ever one type of dimension
$-1$ that is in normal form, the type $\GG\vdash\Hom{\unit}{()}{()}$,
we denote this type $\Obj$ by analogy with the theory $\Glob$. When we
define operations acting on $\Glob_{\unit}$ we always ensure that they
respect the definitional equality by defining them only in normal
form. That means that in practice, we may treat $\Obj$ as a separate
case if needed, keeping in mind that it is simply a short form for
$\Hom{\unit}{()}{()}$.

\paragraph{Semantics.}
We now show that the semantics of the theory $\Glob_{\unit}$ is the
same as the semantics of the theory $\Glob$.
\begin{lemma}\label{lemma:unit-terminal}
  The context $(x:\unit)$ is terminal in the category
  $\Syn{\Glob_{\unit}}$
\end{lemma}
\begin{proof}
  For any context $\GG$ we always have the substitution
  $\GG\vdash\sub{x\mapsto ()}:(x:\unit)$. Moreover, consider a
  substitution $\GG\vdash\Gg:(x:\unit)$.  Then by construction of
  substitutions, $\Gg$ is necessarily of the form
  $\Gg=\sub{x\mapsto t}$ with a derivation for the judgment
  $\GG\vdash t:\unit$.  The conversion rule~$(\eta_{\unit})$ applies
  and gives a definitional equality $\GG\vdash t\equiv
  ():\unit$. Hence, we have a definitional equality between the
  substitutions $\GG\vdash\Gg\equiv\sub{x\mapsto()}:(x:\unit)$. This
  proves that there is a unique morphism $\GG\to(x:\unit)$ in the
  syntactic category $\Syn{\Glob}$.
\end{proof}
\begin{lemma}
  The syntactic categories $\Syn{\Glob}$ and $\Syn{\Glob_{\unit}}$ are
  equivalent.
\end{lemma}
\begin{proof}
  Since the theory $\Glob_{\unit}$ contains all the rules of the
  theory $\Glob$, any valid context $\GG\vdash$ (\resp any valid
  substitution $\GD\vdash \Gg:\GG$) in the theory $\Glob$ also defines
  a valid context $\GG\vdash$ (\resp a valid substitution
  $\GD\vdash \Gg : \GG$) in the theory $\Glob_{\unit}$. Hence, there
  is an inclusion functor $\Syn{\Glob} \to \Syn{\Glob_{\unit}}$. We
  show that this functor is an equivalence, by showing that it is
  fully faithful and essentially surjective. First we show that it is
  faithful by induction on the length of the target context: If the
  target context is empty, then it is terminal, so the statement is
  vacuous, consider two distinct substitutions
  $\GD\vdash \sub{\Gg,x\mapsto t}:(\GG,x:A)$ and
  $\GD\vdash\sub{\Gg',x\mapsto t'}:(\GG,x:A)$ in the theory
  $\Glob$. Then either the two substitutions $\Gg$ and $\Gg'$ are
  distinct in $\Glob$, in which case the induction hypothesis shows
  that they are distinct in $\Glob_{\unit}$, or necessarily the terms
  $t$ and $t'$ are distinct in $\Glob$. Since these terms are in
  $\Glob$, they are not of type $\unit$, so there cannot be a
  definitional equality between them in $\Glob_{\unit}$. This proves
  that $\sub{\Gg,x\mapsto t}$ and $\sub{\Gg',x\mapsto t'}$ define
  distinct arrows in $\Syn{\Glob_{\unit}}$, hence the functor is
  faithful. We show that it is full by considering two contexts $\GD$
  and $\GG$ of $\Glob$ together with a substitution
  $\GD\vdash \Gg:\GG$ in $\Glob_{\unit}$. Then the substitution is
  built out of terms of the context $\GD$, and since none of the
  variables in $\GG$ has type $\unit$, none of those terms have type
  $\unit$, so they are all variables, of $\GD$, and hence the
  substitution $\Gg$ is in fact definable in $\Glob$. Finally, we
  prove that this functor is essentially surjective. Indeed, first
  note that for all context $\GG$, we have the pullback in
  $\Syn{\Glob_{\unit}}$:
  \[
    \begin{tikzcd}
      (\GG,x:\unit) \ar[r]\ar[d]\ar[dr,phantom,"\lrcorner" very near start] & (x:\unit)\ar[d] \\
      \GG\ar[r] & \emptycontext
    \end{tikzcd}
  \]
  Since both $(x:\unit)$ and $\emptycontext$ are terminal objects in
  $\Syn{\Glob_{\unit}}$, the display map $(\GG,x:\unit)\to\GG$ is an
  isomorphism. Using this fact we can recursively eliminate all the
  variables of type $\unit$ in a context and show that every context
  is isomorphic to one without any variable of type $\unit$, that is,
  it is isomorphic to a context of $\Glob$.
\end{proof}

\begin{prop}\label{prop:models-unit}
  There is an equivalence between the models of the theory $\Glob$ and
  the models of the theory $\Glob_{\unit}$.
\end{prop}
\begin{proof}
  First note that the equivalence functor
  $\Syn{\Glob}\to\Syn{\Glob_{\unit}}$ induces an equivalence of
  categories by pre-composition
  $\Set^{\Syn{\Glob_{\unit}}}\to \Set^{\Syn{\Glob}}$.  Since any type
  (\resp any term) in the theory $\Glob$ also defines a type (\resp a
  term) in the theory $\Glob_{\unit}$, the functor
  $\Syn{\Glob}\to \Syn{\Glob_{\unit}}$ can be seen as a functor above
  the category $\Fam$. Moreover, it is straightforward from the
  syntactic definition of this functor that it preserves the terminal
  object and the context extension on the nose, hence this functor
  defines morphism of category with families.This shows that the
  pre-composition restricts to the models to provide a functor
  $\Mod{\Syn{\Glob_{\unit}}} \to \Mod{\Syn{\Glob}}$. Consider the
  functor $f : \Syn{\Glob_{\unit}}\to \Syn{\Glob}$ which is the
  adjoint inverse of the inclusion. Remark that the functor $f$ cannot
  be made into a functor of categories with families, since for
  instance it sends the context $(\emptycontext,x:\unit)$ onto the
  context $\emptycontext$, so it cannot preserve the context extension
  on the nose. However, by definition $f$ preserves all limits, so in
  particular it preserves the terminal objects and the pullbacks along
  display maps. Considering a model $F : \Syn{\Glob} \to \Set$, this
  implies that $f^\star F$ also preserves the terminal object and the
  pullbacks along the display maps, so it is a model and hence
  $f^\star$ restricts as a functor on the models
  $f^\star : \Mod{\Syn{\Glob}} \to \Mod{\Syn{\Glob_{\unit}}}$,
  defining an inverse to the previous functor, hence the equivalence
  of categories.
\end{proof}


\paragraph{The desuspension operation.}
In order to express the theory $\MCaTT$ we need an operation that we
call the \emph{desuspension}, that we first describe as associating,
to every context $\GG$ of $\Glob$, the context $\chop\GG$ in
$\Glob_{\unit}$. It is defined by induction on the raw syntax of the
theory $\Glob$, together with the corresponding operation on types of
$\Glob$
\begin{align*}
  \chop{\emptycontext} &= \emptycontext & \chop{(\GG,x:A)} &= (\chop{\GG} , x:\chop{A})
  &
    \chop{\Obj} &= \unit & \Hom Axy &= \Hom {\chop A}xy
\end{align*}
\begin{prop}
  The desuspension respects the judgments, the following rules are
  derivable {\small
    \[
      \inferrule{\GG\vdash_{\Glob}}{\chop{\GG}\vdash_{\Glob_{\unit}}}
      \qquad\qquad
      \inferrule{\GG\vdash_{\Glob}A}{\chop{\GG}\vdash_{\Glob_{\unit}}
        \chop A} \qquad\qquad
      \inferrule{\GG\vdash_{\Glob}x:A}{\chop{\GG}\vdash_{\Glob_{\unit}}
        x:\chop A}
    \]}
\end{prop}
\begin{proof}
  We prove this result by mutual induction on the derivation tree of
  valid judgments.
  \begin{induction}
  \item \emph{Induction case for contexts}:
    \begin{itemize}
    \item A derivable context obtained by the rule \regle{ec} is
      necessarily the context $\emptycontext\vdash_{\Glob}$. The rule
      \regle{ec} gives a derivation of
      $\chop{\emptycontext} \vdash_{\Glob_{\unit}}$
    \item A derivable context obtained by the rule \regle{ce} is of
      the form $(\GG,x:A)\vdash_{\Glob}$, by the induction case on
      types we have $\chop\GG\vdash_{\Glob_{\unit}} \chop A$. The
      rule~\regle{ce} yields a derivation for
      $\chop\GG,x:\chop A\vdash_{\Glob_{\unit}}$.
    \end{itemize}
  \item \emph{Induction for types}:
    \begin{itemize}
    \item A derivable type obtained by the rule \regle{$\Obj$-intro}
      is of the form $\GG\vdash_{\Glob} \Obj$, we necessarily have
      $\GG\vdash_{\Glob}$. The induction case for contexts implies
      $\chop{\GG}\vdash_{\Glob_{\unit}}$. The rule
      \regle{$\unit$-intro} then provides a derivation for
      $\chop\GG\vdash_{\Glob_{\unit}}\unit$.
    \item A derivable type obtained by the rule
      \regle{$\Hom{}{}{}$-intro} is of the form
      $\GG\vdash_{\Glob} \Hom Atu$. The induction cases for types and
      variables provide derivations for
      $\chop\GG\vdash_{\Glob_{\unit}} A$,
      $\chop\GG\vdash_{\Glob_{\unit}} t:\chop A$ and
      $\chop\GG\vdash_{\Glob_{\unit}} u:\chop A$. The rule
      \regle{$\Hom{}{}{}$-intro} then gives a derivation for
      $\chop\GG\vdash \Hom{\chop A}tu$.
    \end{itemize}
  \item \emph{Induction for variables}: A term obtained by the rule
    \regle{Var} is a variable $\GG\vdash_{\Glob} x:A$, and we have
    $\GG\vdash_{\Glob}$. The induction for contexts implies
    $\chop\GG\vdash_{Glob_{\unit}}$. Moreover, $(x:A)\in\GG$ implies
    $(x:\chop A)\in\chop\GG$. The rule~\regle{var} lets us construct a
    derivation for $\chop\GG\vdash_{\Glob_{\unit}} x:\chop
    A$. \qedhere
  \end{induction}
\end{proof}
\noindent From now on, when we perform proofs by induction on the
derivation tree, we rely on the form of the judgment, for instance, we
may write: ``For the context $\emptycontext\vdash$'' to mean that we
discriminate on the rule \regle{ec} and that in this case the context
is necessarily $\emptycontext$.  This is justified since the rule
\regle{ec} is the only one that allows to derive
$\emptycontext\vdash$, and more generally each syntactic
constructor constructor corresponds to exactly one introduction rule.

\paragraph{Examples}
We illustrate the desuspension with a few examples. Intuitively merely
consists in rewriting the type $\Obj$ into the type $\unit$. For each
of the context in $\Glob_{\unit}$ we also give a simpler context to
which it is isomorphic.
\[
  \begin{array}{c|c}
    \GG & \chop\GG \\
    \hline
    (x:\Obj) & (x:\unit)\simeq\emptycontext \\
    (x:\Obj,y:\Obj,f:\Hom{\Obj}xy) & (x:\unit,y:\unit,f:\Hom{\unit}xy)\simeq
                                     (f:\Obj)\\
    (x:\Obj,y:\Obj,f:\Hom{\Obj}xy,g:\Hom{\Obj}yx) & (x:\unit,y:\unit,f:\Hom{\unit}xy,g:\Hom{\unit}zy)\simeq
                                                    (f:\Obj,g:\Obj)\\
  \end{array}
\]

\subsection{The theory $\MCaTT$}
The type theory $\MCaTT$ relies on the theory $\CaTT$ on a fundamental
level: Its inference rules use the derivability of judgments in
$\CaTT$. To express these rules we need to generalize the desuspension
as an operation from the raw syntax of $\CaTT$ to the raw syntax of
$\MCaTT$. The raw syntax of $\MCaTT$ is obtained from the one of
$\Glob_{\unit}$ by adding two term constructors $\mop$ and $\mcoh$.  A
term expression is either a variable, or of the form
$\mop_{\GG,A}[\Gg]$ or $\mcoh_{\GG,A}[\GG]$ with $\GG$ a ps-context
and $A$ a type and $\Gd$ a substitution. We sometimes unite these two
cases under the notation $\mconstr$ which means either one of $\mop$
or $mcoh$. The variable set of these terms is not well defined (since
not invariant under definitional equality), but the action of
substitution is, and it is defined by
$\mconstr_{\GG,A}[\Gg][\Gd] = \mconstr_{\GG,A}[\Gg\circ\Gd]$. By
convention, whenever we use $\constr$ and $\mconstr$ in the same
equality, they are corresponding constructors, so either $\cohop$ and
$\mop$ or $\coh$ and $\mcoh$.


\paragraph{The general desuspension operation.}
We generalize the desuspension operation to contexts, types, terms and
substitutions expressions of the theory $\CaTT$ as follows
\begin{align*}
  \chop\emptycontext &= \emptycontext & \chop{(\GG,x:A)} &=
                                                           \chop{\GG} , x:\chop{A}\\
  \chop{\Obj} &= \unit & \chop{(\Hom Atu)} &= \Hom{\chop A}{\chop t}{\chop u} \\
  \chop x &= x &  \chop{\constr_{\GG,A}[\Gg]} &= \mconstr_{\GG,A}[\chop \Gg]\\
  \chop{\sub{}} &= \sub{} & \chop{\sub{\Gg,x\mapsto t}} &=
                                                          \sub{\chop \Gg, x\mapsto \chop t}
\end{align*}

\paragraph{The theory $\MCaTT$.}
The introduction rules for the constructors $\mop$ and $\mcoh$ reuse a
lot of the machinery that we have developed for the theory $\CaTT$. We
use the desuspension operation to transfer this machinery to the
theory $\MCaTT$.  \small{
  \[
    \inferrule{\GG\vdashps \\ \GG\vdashop A \\ \GD\vdash \Gg :
      \chop{\GG}}{\GD\vdash\mop_{\GG,A} :
      (\chop{A})[\Gg]}{\regle{$\mop$-intro}} \qquad
    \inferrule{\GG\vdashps \\ \GG\vdasheq A~\\ \GD\vdash
      \Gg:\chop{\GG}}
    {\GD\vdash\mcoh_{\GG,A}:(\chop{A})[\Gg]}{\regle{$\mcoh$-intro}}
  \]
} In the above rules, the judgments $\GG\vdashps$, $\GG\vdashop A$ and
$\GG\vdasheq A$ are the judgments defined for the theory $\CaTT$. In
particular, although $A$ is used as an index for the terms of the
theory $\MCaTT$, it may not be a valid type in this theory, but it has
to be a valid type in the theory $\CaTT$. For instance in the second of
our following example, the type $A$ uses the expression
$\texttt{comp}$ that we have defined as a constructor of $\CaTT$ and
not of $\MCaTT$. All the rules of the theory $\MCaTT$ are summarized
in Appendix~\ref{app:mcatt} to provide a self-contained
overview.




\paragraph{Examples.} We do not provide an implementation for this
theory, but we can check by hand a few examples of derivations as a
sanity check for our definition of monoidal weak $\omega$-categories.
\begin{itemize}
\item Monoidal product: First consider the ps-context
  $\GG_{\texttt{comp}}$, that we have defined in
  Section~\ref{sec:catt}.  We have
  $\chop{\GG_{\texttt{comp}}} \simeq (f:\Obj,g:\Obj)$, thus for any
  context $\GD$, a pair of terms $t,u$ of type $\Obj$ in $\GD$ define
  a unique substitution $\GD\vdash \Gg : \GG_{\texttt{comp}}$. The
  rule~\regle{$\mop$-intro} gives a derivation of the product of $t$
  and $u$ as
  $\GD\vdash \mop_{\GG_{\texttt{comp},\Hom{} xz}} [\Gg]: \Obj$. We
  simplify this notation as $\GD\vdash\texttt{prod}\ t\ u : \Obj$.
\item Associativity of monoidal product: Similarly, consider the
  context $\Gamma_{\texttt{assoc}}$ defined in Section~\ref{sec:catt}.
  A context $\GD$ in $\MCaTT$ equipped with three terms $t,u,v$ of
  type $\Obj$ define a unique substitution
  $\GD\vdash \Gg:\chop\GG_{\texttt{assoc}}$. Hence the
  rule~\regle{$\mcoh$-intro} provides an associator for the monoidal
  product
  $\GD\vdash \mop_{\GG_{\texttt{assoc}},\Hom{}{\comp\ (\comp\ f\ g)\ h
    }{\comp\ f (\comp\ g\ h)}} : \Hom{}{\texttt{prod}\ (\texttt{prod}\
    t\ u )\ v}{\texttt{prod}\ t\ (\texttt{prod}\ u\ v)}$ We simplify
  this notation as
  $\GD\vdash \texttt{passoc}\ t\ u\ v: \Hom{}{\texttt{prod}\
    (\texttt{prod}\ t\ u )\ v}{\texttt{prod}\ t\ (\texttt{prod}\ u\
    v)}$.
\item Neutral element: Consider the ps-context
  $\GG_{\texttt{neutral}}=(x:\Obj)$. Then
  $\chop{\GG_{\texttt{neutral}}}$ is terminal. The
  rule~\regle{\mcoh-intro} allows to derive the unit in any context
  $\Delta$ as
  $\GD\vdash \mcoh_{\GG_{\texttt{neutral}},\Hom{}xx} : \Obj$.
\end{itemize}

\section{Syntactic categories of $\CaTT$ and $\MCaTT$}
We now relate the two theories $\CaTT$ and $\MCaTT$ to each other via
a pair of functorial translations. We have already seen one of these
translations: the desuspension, the other one is called the
\emph{reduced suspension}. Our approach for defining these translations
follows the same plan: first define them on a raw syntactic level, and then
lift them to the syntactic categories by showing that they preserve the judgments.
We then prove that these two translations are closely related, as they define a coreflective
adjunction between the syntactic categories.

\subsection{The desuspension functor}
We have already defined the desuspension operation, that we have used
freely on the raw syntax of the theory. We now lift this operation to
the syntactic categories by showing that it respects the derivability
of judgments. We first prove following result about the interaction of
the desuspension with the application of substitutions on a raw syntax
level.
\begin{lemma}\label{lemma:syn-desusp}
  Given a substitution $\GD\vdash \Gg:\GG$, for any type
  $\GG\vdash A$, we have $\chop{(A[\Gg])} = \chop A[\chop \Gg]$, for
  any term $\GG\vdash t:A$, we have
  $\chop{(t[\Gg])} = \chop t[\chop\Gg]$ and for any substitution
  $\GG\vdash\Gth : \GTH$, we have
  $\chop{(\Gth\circ \Gg)} = \chop\Gth\circ \chop\Gg$.
\end{lemma}
\begin{proof}
  We prove this by mutual induction
  \begin{induction}
  \item \emph{Induction for types}:
    \begin{itemize}
    \item In the case of the type $\Obj$, we have
      $\chop{(\Obj[\Gg])} = \unit$ and
      $\chop{\Obj}[\chop\Gg] = \unit$.
    \item In the case of a type of the form $\Hom{A} tu$, we have the
      following equalities
      \begin{align*}
        \chop{((\Hom Atu)[\Gg])} &= \Hom{\chop(A[\Gg])}{\chop{(t[\Gg])}}{\chop{(u[\Gg])}} &
                                                                                            \chop{(\Hom{A}{t}{u})}[\chop\Gg] &= \Hom{\chop A[\chop\Gg]}{\chop t[\chop\Gg]}{\chop u[\chop\Gg]}
      \end{align*}
      and we conclude by induction on types and terms.
    \end{itemize}
  \item \emph{Induction for terms}:
    \begin{itemize}
    \item In the case of a variable $\GG\vdash x:A$, denote
      $t = x[\Gg]$. Then the association $x\mapsto \chop{t}$ appears
      in $\chop\Gg$. Hence $x[\chop\Gg] = \chop{(x[\Gg])}$.
    \item In the case of a term of the form $\constr_{\GG,A}[\Gd]$,
      the following equalities hold
      \begin{align*}
        \chop{(\constr_{\GG,A}[\Gd][\Gg])} &= \mconstr_{\GG,A}[\chop{(\Gd\circ\Gg)}] &
                                                                                       \chop{(\constr_{\GG,A}[\Gd])}[\chop\Gg] &= \mconstr_{\GG,A}[\chop\Gd\circ\chop\Gg]
      \end{align*}
      The induction case for substitution then provides the equality
      between these two expressions.
    \end{itemize}
  \item \emph{Induction for substitutions}:
    \begin{itemize}
    \item In the case of the empty substitution $\sub{}$, we have
      $\sub{}\circ\Gg = \sub{}$, and hence
      $\chop{\pa{\sub{}\circ \Gg}} = \sub{}$. But we also have
      $\chop{\sub{}}\circ \Gg = \sub{}$.
    \item In the case of a substitution of the form
      $\sub{\Gth,x\mapsto t}$, we have the following equalities
      \begin{align*}
        \chop{(\sub{\Gth,x\mapsto t}\circ\Gg)} &= \sub{\chop{(\Gth\circ\Gg)},x\mapsto \chop{(t[\Gg])}} &
                                                                                                         \chop{\sub{\Gth,x\mapsto t}}\circ\chop{\Gg} &= \sub{\chop{\Gth}\circ\chop\Gg,x\mapsto\chop t [\chop \Gg]}
      \end{align*}
      Then the induction case for substitutions proves that
      $\chop{(\Gth\circ\Gg)} = \chop\Gth\circ\chop\Gg$, and the
      induction case for terms applies and shows that
      $\chop{(t[\Gg])} = \chop t[\chop \Gg]$.  This proves the
      equality between the two above expressions. \qedhere
    \end{itemize}
  \end{induction}
\end{proof}

\begin{prop}\label{prop:desusp-correct}
  The rules are admissible {\small
    \[
      \inferrule{\GG\vdash_{\CaTT}}{\chop{\GG}\vdash_{\MCaTT}}
      \qquad\qquad
      \inferrule{\GG\vdash_{\CaTT}A}{\chop{\GG}\vdash_{\MCaTT}\chop A}
      \qquad\qquad
      \inferrule{\GG\vdash_{\CaTT}t:A}{\chop{\GG}\vdash_{\MCaTT}\chop
        t : \chop A} \qquad\qquad \inferrule{\GD\vdash_{\CaTT}
        \Gg:\GG}{\chop{\GD}\vdash_{\MCaTT} \chop\Gg: \chop\GG}
    \]}
\end{prop}
\begin{proof}
  We prove this result by mutual induction on the derivation of the
  judgements.
  \begin{induction}
  \item \emph{Induction for contexts}:
    \begin{itemize}
    \item For the empty context $\emptycontext\vdash_{\CaTT}$, we have
      $\chop{\emptycontext} = \emptycontext$ and the rule~\regle{ec}
      gives a derivation of $\chop{\emptycontext}\vdash_{\MCaTT}$.
    \item For the context $(\GG,x:A)\vdash_{\CaTT}$, we necessarily
      have $\GG\vdash_{\CaTT}$ and $\GG\vdash_{\CaTT} A$. By
      induction, we have $\chop\GG\vdash_{\MCaTT}$ and
      $\chop{\GG}\vdash_{\MCaTT} \chop A$. Hence the rule~\regle{ce}
      gives a derivation for $(\chop{\GG},x:\chop A)\vdash$. We
      conclude by noticing that
      $\chop{(\GG,x:A)} = (\chop\GG,x:\chop A)$.
    \end{itemize}
  \item \emph{Induction for types}:
    \begin{itemize}
    \item For the type $\GG\vdash_{\CaTT}\Obj$, we necessarily have
      $\GG\vdash_{\CaTT}$. By induction, this implies
      $\chop{\GG}\vdash_{\MCaTT}$. The rule~\regle{$\unit$-intro}
      gives a derivation for $\chop\GG\vdash\chop\Obj$.
    \item For the type $\GG\vdash_{\CaTT} \Hom Atu$ we have a
      derivation of $\GG\vdash_{\CaTT} A$, $\GG\vdash_{\CaTT}t:A$ and
      $\GG\vdash_{\CaTT}u:A$. By induction, we get a derivation of
      $\chop\GG\vdash_{\MCaTT}\chop A$,
      $\chop \GG\vdash_{\MCaTT} \chop t: \chop A$ and
      $\chop \GG\vdash_{\MCaTT}\chop u : \chop A$. The
      rule~\regle{$\Hom{}{}{}$-intro} then provides a derivation of
      $\chop\GG\vdash\Hom{\chop A}{\chop t}{\chop u}$.
    \end{itemize}
  \item \emph{Induction for terms}:
    \begin{itemize}
    \item For a variable $\GG\vdash_{\CaTT} x:A$, we necessarily have
      $\GG\vdash_{\CaTT}$. By induction, it provides a derivation of
      $\chop\GG\vdash_{\MCaTT}$. Moreover, we have the condition
      $(x:A)\in\GG$, hence $(x:\chop A)\in\chop \GG$. The
      rule~\regle{var} then proves $\chop\GG\vdash x:\chop A$.
    \item For a term of the form
      $\GD\vdash_{\CaTT}\constr_{\GG,A}[\Gg]:A[\Gg]$, we have a
      derivation of $\GD\vdash_{\CaTT}\Gg:\GG$. By induction, provides
      a derivation for $\chop\GD\vdash_{\MCaTT}
      \chop\Gg:\chop\GG$. The rule~\regle{$\mop$-intro}
      or~\regle{\mcoh-intro} then gives a derivation of
      $\chop\GD\vdash_{\MCaTT} \mconstr_{\GG,A}[\chop
      \Gg]:\chop{A}[\chop\Gg]$.
    \end{itemize}
  \item \emph{Induction for substitutions}:
    \begin{itemize}
    \item For the empty substitution
      $\GD\vdash_{\CaTT}\sub{}:\emptycontext$, we have a derivation of
      $\GD\vdash_{\CaTT}$. By induction, it provides a derivation of
      $\chop\GD\vdash_{\MCaTT}$. The rule~\regle{es} then proves
      $\chop\GD\vdash_{\MCaTT}\sub{}:\emptycontext$.
    \item For the substitution
      $\GD\vdash_{\CaTT}\sub{\Gg,x\mapsto t}:(\GG,x:A)$, we have
      derivations of $\GD\vdash_{\CaTT}\Gg:\GG$,
      $\GG,x:A\vdash_{\CaTT}$ and $\GD\vdash_{\CaTT}t:A[\Gg]$.  By
      induction those provide derivations of
      $\chop\GD\vdash_{\MCaTT} \chop\Gg:\chop\GG$,
      $\chop{\GG},x:\chop A \vdash_{\MCaTT}$ and
      $\chop{\GD}\vdash_{\MCaTT} \chop t : \chop{(A[\Gg])}$. Moreover,
      by Lemma~\ref{lemma:syn-desusp}, the last judgment rewrites as
      $\chop\GD \vdash \chop t: \chop A[\chop\Gg]$. Hence the
      rule~\regle{es} provides a derivation of the judgment
      $\chop\GD\vdash_{\MCaTT}\sub{\chop\Gg , x\mapsto \chop t} :
      (\chop \GG,x:\chop A)$.
    \end{itemize}
  \end{induction}
\end{proof}

\paragraph{Categorical reformulation.}
We reformulate Lemma~\ref{lemma:syn-desusp} and
Proposition~\ref{prop:desusp-correct} in a categorical fashion, taking
advantage of the formalism of categories with families.
\begin{coro}\label{coro:desusp-cwf-mor}
  The desuspension defines a morphism of categories with families
  $\chop{}:\Syn{\CaTT} \to \Syn{\MCaTT}$.
\end{coro}
\begin{proof}
  We have proved in Proposition~\ref{prop:desusp-correct} that the
  desuspension sends contexts (\resp substitutions) in $\CaTT$ onto
  contexts (\resp substitutions) in $\MCaTT$, and
  Lemma~\ref{lemma:syn-desusp} shows that it respects
  composition. Hence, to prove that it is a functor, it suffices to
  show that it preserves identity, which we do by induction.
  \begin{itemize}
  \item For the empty context $\emptycontext$, the identity is the
    empty substitution $\id\emptycontext = \sub{}$, and we have
    $\chop{\sub{}} = \sub{}$.
  \item For the context $(\GG,x:A)$, the identity is
    $\sub{\id\GG,x\mapsto x}$, whose image is
    $\sub{\chop{\id\GG},x\mapsto x}$.  By induction, we have
    $\chop{\id\GG} = \id{\chop\GG}$, which gives the equality
    $\chop{\id{(\GG,x:A)}} = \id{\chop{(\GG,x:A)}}$.
  \end{itemize}
  We now show that this functor is a morphism of categories with
  families, be defining the image of a type $\GG\vdash A$ to be
  $\chop A$ and the image of a term $\GG\vdash t:A$ to be $\chop
  t$. Proposition~\ref{prop:desusp-correct} shows that these elements
  live in the adequate set, and Lemma~\ref{lemma:syn-desusp} moreover
  ensures that this association is functorial, so $\chop{}$ defines a
  morphism in the slice category $\Cat/\Fam$. Since moreover the
  operation $\chop{}$ is defined to preserve the terminal context and
  the context comprehension (by definition, we have
  $\chop\emptycontext = \emptycontext$ and
  $\chop{(\GG,x:A)} = (\chop\GG,x:\chop A)$ and
  $\chop{\sub{\Gg,x\mapsto t}} = \sub{\chop\Gg,x\mapsto \chop t}$), it
  is a morphism of categories with families.
\end{proof}


\subsection{The reduced suspension functor}
We define another translation going in the opposite direction called
the \emph{reduced suspension}. It is an operation associating an
expression of the theory $\CaTT$ to every expression of the theory
$\MCaTT$. This operation is not however purely syntactic, and even
though it outputs a raw expression of $\CaTT$, it is only properly
defined on the derivable expression of $\MCaTT$.

\paragraph{The reduced suspension operation.}
In order to define the reduced suspension, we assume the existence of
a variable name that is completely fresh, and call it $\new{}$. This
could be achieved by extending the set of variables that we use with
the variable $\new{}$.  
%
\[
  \begin{array}{rr@{\ =\ }l@{\qquad\qquad}rr@{\ =\ }l}
    \emph{judgment} & \multicolumn{2}{c}{\emph{definition}} &  \emph{judgment} & \multicolumn{2}{c}{\emph{definition}}\\
    \emptycontext\vdash
                    & \deloop\emptycontext & (\new {}:\Obj)
                                                                               & \GG,x:\unit\vdash
                    & \deloop(\GG,x:\unit) & \deloop\GG\\
                    & \multicolumn{2}{c}{} & \GG,x:A\vdash
                                                                               & \deloop(\GG,x:A) & (\deloop\GG ,x:\deloop A) \\[1.5em]
    \GG\vdash\unit & \deloop\unit & \Obj & \GG\vdash \Obj & \deloop\Obj & \Hom{\Obj}{\new{}}{\new{}} \\
                    & \multicolumn{2}{c}{} & \GG\vdash\Hom Atu
                                                                               & \deloop{\pa{\Hom Atu}} & \Hom{\deloop A}{\deloop t}{\deloop u}\\[1.5em]
    \GG\vdash \texttt{()}:\unit & \deloop{\texttt{()}} & \new{}
                                                                               & \GG\vdash x:A\ (A\neq\unit) &  \deloop x & x \\
                    & \multicolumn{2}{c}{} & \GG\vdash\mconstr_{\GTH,A}[\Gg] : A[\Gg]
                                                                               & \deloop{\mconstr_{\GTH,A}[\Gg]} & \constr_{\GTH,A}[\red\GTH\circ\deloop\Gg] \\[1.5em]
    \GG\vdash \sub{}:\emptycontext & \deloop{\sub{}} & \sub{\new{}\mapsto \new{}}
                                                                               & \GG\vdash\sub{\Gg,x\mapsto  t}:(\GD,x:\unit) & \deloop{\sub{\Gg,x\mapsto t}} & \deloop\Gg\\
                    & \multicolumn{2}{c}{}
                                                            &\GG\vdash\sub{\Gg,x\mapsto t}:(\GD,x:A)
                                                                               & \deloop{\sub{\Gg,x\mapsto t}} & \sub{\deloop\Gg,x\mapsto \deloop t}\\[1.5em]
    \multicolumn{6}{l}{\text{Where the substitution $\red\GTH$ is defined by induction
    on the context $\GTH\vdash$  of $\CaTT$ by}}\\
                    &  \red{\emptycontext} & \sub{} & & \red{(\GTH,x:A)} & \sub{\red\GTH,x\mapsto \deloop{\chop x}}
  \end{array}
\]
In the case of a variable $\GG\vdash x:A$ we have to assume that
$A\neq\unit$ to ensure that the term is in normal form so that
definitional equality is respected.
\begin{rem}
  We do not define the reduced suspension as a function on the raw
  syntax of $\MCaTT$ because we rely on the normal form in the theory
  $\MCaTT$, which is not defined on the raw syntax.  To define the
  reduced suspension of a variable term $x$, we need to know whether
  $\GG\vdash x:\unit$ is derivable, in which case the term is sent
  onto $\bullet$.
\end{rem}

\paragraph{Syntactic properties of the reduced suspension.}
We show a few technical results, that are useful to prove that the
reduced suspension preserves the judgments.
\begin{lemma}\label{lemma:rsusp-preserves-new}
  For all substitution $\GD\vdash_{\MCaTT}\Gg:\GG$, we have
  $\new {}[\deloop\Gg] = \new {}$.
\end{lemma}
\begin{proof}
  By definition of $\deloop{\Gg}$, the only mapping $\new{}\mapsto t$
  in $\deloop\Gg$ is $\new {}\mapsto\new {}$.
\end{proof}

\begin{lemma}\label{lemma:syn-rsusp}
  Given a substitution $\GD\vdash_{\MCaTT}\Gg:\GG$, the following hold
  \begin{itemize}
  \item for any type $\GG\vdash_{\MCaTT} A$ we have the equality
    $\deloop{(A[\Gg])}=\deloop A[\deloop \Gg]$
  \item for any term $\GG\vdash_{\MCaTT} t:A$, we have the equality
    $\deloop{(t[\Gg])} = \deloop t[\deloop \Gg]$
  \item for any substitution $\GG\vdash_{\MCaTT}\Gd:\GD$, we have the
    equality $\deloop{\Gd\circ \Gg} = \deloop\Gd\circ\deloop\Gg$.
  \end{itemize}
\end{lemma}
\begin{proof}
  We suppose given the substitution $\Gg$ and prove these three
  results by mutual induction
  \begin{induction}
  \item \emph{Induction for types}:
    \begin{itemize}
    \item For the type $\unit$, we have $\unit[\Gg] = \unit$, hence
      $\deloop{\pa{\unit[\Gg]}} = \deloop \unit = \Obj$. But we also
      have $\deloop\unit[\deloop \Gg] = \Obj[\deloop\Gg]=\Obj$.
    \item For the type $\Hom Atu$, we have the two following
      equalities
      \begin{align*}
        \deloop{((\Hom Atu)[\Gg])} &= \Hom{\deloop{(A[\Gg])}}{\deloop{(t[\Gg])}}{\deloop{(u[\Gg])}} &
                                                                                                      (\deloop{(\Hom Atu)})[\deloop\Gg] &= \Hom{(\deloop A)[\deloop \Gg]}{(\deloop t)[\deloop \Gg]}{(\deloop u)[\deloop \Gg]}
      \end{align*}
      and by induction, we have
      $\deloop{(A[\Gg])} = (\deloop A)[\deloop\Gg]$,
      $\deloop{(t[\Gg])} = (\deloop t)[\deloop \Gg]$ and
      $\deloop{(u[\Gg])} = (\deloop u)[\deloop \Gg]$.
    \end{itemize}
  \item \emph{Induction for terms}:
    \begin{itemize}
    \item For the term $\texttt{()}$, we have
      $\deloop{\pa{\texttt{()}[\Gg]}} = \new{}$. and also
      $\deloop{\texttt{()}}[\deloop \Gg] = \new{}[\deloop\Gg]$.
      Lemma~\ref{lemma:rsusp-preserves-new} shows
      $\deloop{\texttt{()}}[\deloop\Gg] = \new{}$.
    \item For the term $\GG\vdash x:A$ ($\neq \unit$), by definition,
      $\deloop\Gg$ defines the mapping $x\mapsto \deloop{(x[\Gg])}$,
      so $x[\deloop \Gg] = \deloop {(x[\Gg])}$.
    \item For a term of the form $\mop_{\GG,A}[\Gd]$ we have the
      following equalities
      \begin{align*}
        \deloop{(\mconstr_{\GG,A}[\Gd][\Gg])} &= \constr_{\GG,A}[\red\GG\circ\deloop{(\Gd\circ\Gg)}] &
                                                                                                       (\deloop{\mconstr_{\GG,A}[\Gd]})[\deloop \Gg] &= \constr_{\GG,A}[(\red\GG\circ\deloop\Gd)\circ\deloop\Gg]
      \end{align*}
      By induction and associativity of composition, these two terms
      are equal.
    \end{itemize}
  \item \emph{Induction for substitutions}:
    \begin{itemize}
    \item For the substitution $\sub{}$, we have
      $\deloop{(\sub{}\circ\Gg)} = \sub{\new {}\mapsto\new {}}$ and
      $\deloop{\sub{}}\circ\deloop{\Gg} = \sub{\new {}\mapsto\new
        {}[\deloop\Gg]}$.  Lemma~\ref{lemma:rsusp-preserves-new} this
      shows that
      $\deloop{\sub{}}\circ\deloop\Gg = \sub{\new {}\mapsto\new {}}$.
    \item For the substitution
      $\GG\vdash\sub{\Gth,x\mapsto t}:(\GTH,x:\unit)$, we have the
      following equalities
      \begin{align*}
        \deloop{(\sub{\Gth,x\mapsto t}\circ\Gg)} &= \sub{\deloop{(\Gth\circ\Gg)}} &
                                                                                    \deloop{\sub{\Gth,x\mapsto t}}\circ\deloop{\Gg} &= \sub{\deloop\Gth\circ\deloop\Gg}
      \end{align*}
      and we conclude by the induction case for substitutions.
    \item For a substitution of the form
      $\GG\vdash\sub{\Gth,x\mapsto t}:(\GTH,x:A)$ with $A\neq\unit$,
      we have the equalities
      \begin{align*}
        \deloop{(\sub{\Gth,x\mapsto t}\circ\Gg)} &= \sub{\deloop{(\Gth\circ\Gg)},x\mapsto\deloop{(t[\Gg])}} &
                                                                                                              \deloop{\sub{\Gth,x\mapsto t}}\circ\deloop{\Gg} &= \sub{\deloop\Gth\circ\deloop\Gg,x\mapsto(\deloop t)[\deloop\Gg]}
      \end{align*}
      and by conclude by the induction cases for substitutions and terms. \qedhere
    \end{itemize}
  \end{induction}
\end{proof}

\paragraph{Reduced suspension on the theory $\Glob$.}
Note that the theory $\Glob$ can be included in the theory $\MCaTT$,
by sending any expression in the theory $\Glob$ to the same
expression, seen as an expression of $\MCaTT$. The only difference is
that the type $\Obj$ is primitive in $\Glob$ while it is interpreted
as $\Hom{}{\texttt{()}}{\texttt{()}}$ in $\MCaTT$. We first focus on
the restriction of the reduced suspension to the theory $\Glob$.
Since the reduced suspension sends variables on variables, it
preserves the expressions of $\Glob$. We show that this transformation
respects the derivation. This is a technical intermediate for
generalizing the reduced suspension as a transformation from $\MCaTT$
to $\CaTT$.
\begin{lemma}\label{lemma:rsusp-correct-gsett}
  In the theory $\Glob$, the following rules are admissible {\small
    \[
      \inferrule{\GG\vdash_{\Glob}}{\deloop\GG\vdash_{\Glob}}\qquad\qquad
      \inferrule{\GG\vdash_{\Glob} A}{\deloop\GG\vdash_{\Glob} \deloop
        A}\qquad\qquad \inferrule{\GG\vdash_{\Glob}
        t:A}{\deloop\GG\vdash_{\Glob}\deloop t: \deloop A}\qquad\qquad
      \inferrule{\GD\vdash_{\Glob}\Gg:\GG}{\deloop\GD\vdash_{\Glob}\deloop\Gg:\deloop\GG}
    \]}
\end{lemma}
\begin{proof}
  We prove this result by mutual induction on contexts, types, terms
  and substitutions
  \begin{induction}
  \item \emph{Induction for contexts}:
    \begin{itemize}
    \item For the empty context $\emptycontext$, we have
      $\deloop{\emptycontext} = (\new {}:\Obj)$, and we can construct
      a derivation of $\deloop{\emptycontext}\vdash$ by applying
      successively the rules \regle{ec},\regle{$\Obj$-intro} and
      \regle{ce}.
    \item For the context $(\GG,x:A)\vdash$ ($A\neq\unit$ since we are
      in the theory $\Glob$), we have $\GG\vdash A$. By induction this
      gives a derivation for $\deloop\GG\vdash \deloop A$. The
      rule~\regle{ce}, provides a derivation for
      $(\deloop\GG,x:\deloop A)\vdash$
    \end{itemize}
  \item \emph{Induction for types}:
    \begin{itemize}
    \item For the type $\GG\vdash \Obj$, we have a derivation of
      $\GG\vdash$. By the induction, this shows
      $\deloop\GG\vdash$. Since $(\new {}:\Obj)\in\deloop\GG$, we can
      construct a derivation of $\deloop\GG\vdash\deloop\Obj$ as
      follows {\small
        \[
          \inferrule{
            \inferrule{\deloop\GG\vdash}{\deloop\GG\vdash\Obj}{\regle{$\Obj$-intro}} \\
            \inferrule{\deloop \GG\vdash \\
              (\new {}:\Obj)\in\deloop\GG}{\deloop\GG\vdash\new {}:\Obj}{\regle{var}}\\
            \inferrule{\deloop \GG\vdash \\
              (\new {}:\Obj)\in\deloop\GG}{\deloop\GG\vdash\new
              {}:\Obj}{\regle{var}} }{\deloop\GG\vdash \Hom\Obj{\new
              {}}{\new {}}} {\regle{$\Hom{}{}{}$-intro}}
        \]}
    \item For the type $\GG\vdash \Hom Atu$, we have derivations of
      $\GG\vdash A$, $\GG\vdash t:A$ and $\GG\vdash u:A$. By
      induction, those give derivations of
      $\deloop\GG\vdash\deloop A$,
      $\deloop\GG\vdash \deloop t : \deloop A$ and
      $\deloop \GG\vdash \deloop u : \deloop A$.  The
      rule~\regle{$\Hom{}{}{}$-intro} then provides a derivation of
      $\deloop\GG\vdash \Hom{\deloop A}{\deloop t}{\deloop u}$.
    \end{itemize}
  \item \emph{Induction for terms}: A term in $\Glob$ is a variable
    $\GG\vdash x:A$. For such a variable, we have a derivation of
    $\GG\vdash$, and by induction it gives a derivation of
    $\deloop\GG\vdash$. Moreover, the condition $(x:A)\in\GG$ is
    satisfied, and since we are the in the theory $\Glob$,
    $A \neq \unit$. So we have $(x:\deloop A)\in\deloop\GG$. The
    rule~\regle{var} then yields a derivation of
    $\deloop\GG\vdash x:\deloop A$.
  \item \emph{Induction for substitutions}:
    \begin{itemize}
    \item For the substitution $\GD\vdash\sub{}:\emptycontext$, we
      have a derivation of $\GD\vdash$. By induction, this gives a
      derivation of $\deloop\GD\vdash$. Moreover, we also have by
      definition that $(\new {}:\Obj)\in\deloop\GD$, and we have
      already constructed a derivation of $(\new{}:\Obj)\vdash$. This
      lets us construct a derivation for
      $\deloop\GD \vdash \deloop{\sub{}}:\deloop{\emptycontext}$ as
      follows {\small
        \[
          \inferrule*[Right=\regle{se}]{
            \inferrule*[Right=\regle{es}]{\deloop\GD\vdash}{\deloop{\GD}\vdash\sub{}:\emptycontext}\\
            (\new {}:\Obj)\vdash \\
            \inferrule*[Right=\regle{var}]{\deloop\GD\vdash \\
              (\new {}:\Obj)\in\deloop\GD}{\deloop\GD\vdash\new
              {}:\Obj}} {\deloop\GD\vdash\sub{\new {}\mapsto\new {}} :
            (\new {}:\Obj)}
        \]}
    \item For the substitution
      $\GD\vdash\sub{\Gg,x\mapsto t}:(\GG,x:A)$, we have $A \neq\unit$
      since we are in $\Glob$. Then we necessarily have derivations of
      $\GD\vdash \Gg:\GG$, $(\GG,x:A)\vdash$ and $\GD\vdash t:A[\Gg]$.
      By induction, those give derivations of the judgments
      $\deloop\GD\vdash\deloop\Gg:\deloop\GG$,
      $\deloop\GG,x:\deloop A\vdash$ and
      $\deloop{\GD}\vdash \deloop t:
      \deloop{A[\Gg]}$. Lemma~\ref{lemma:syn-rsusp} allows to rewrite
      the last of these judgments as
      $\deloop\GD\vdash \deloop t:\deloop A[\deloop \Gg]$. The
      rule~\regle{se} then provides a derivation of
      $\deloop\GD\vdash\deloop{\sub{\Gg,x\mapsto
          t}}:\deloop{(\GG,x:A)}$. \qedhere
    \end{itemize}
  \end{induction}
\end{proof}

\paragraph{Properties of the substitution $\red\GD$.}
The definition of the reduced suspension relies on the substitution
$\red\GD$. We prove syntactic properties about this substitution.
\begin{lemma}\label{lemma:syn-red}
  We have the following result for the action of the substitution
  $\red\GD$ on terms, types and substitutions.
  \begin{itemize}
  \item For any type $\GD\vdash_{\CaTT} A$, we have
    $A[\red\GD] = \deloop{\chop A}$
  \item For any term $\GD\vdash_{\CaTT} t:A$, we have
    $t[\red\GD] = \deloop{\chop t}$.
  \item For any substitution $\GD\vdash_{\CaTT} \Gg:\GG$, we have
    $\Gg\circ\red\GD = \red\GG\circ\deloop{\chop \Gg}$.
  \end{itemize}
\end{lemma}
\begin{proof}
  We prove these equalities by mutual induction
  \begin{induction}
  \item \emph{Induction on types}:
    \begin{itemize}
    \item For the type $\GD\vdash_{\CaTT}\Obj$, we have
      $\deloop{\chop\Obj} = \Obj$ and by definition,
      $\Obj[\red\GD] = \Obj$.
    \item For the type $\GD\vdash_{\CaTT}\Hom Atu$, we have the
      equalities
      \begin{align*}
        (\Hom Atu)[\bullet_\GD] &= \Hom {A[\bullet_\GD]}{t[\bullet_\GD]}{u[\bullet_\GD]} &
                                                                                           \deloop{\chop{(\Hom Atu)}} &= \Hom{\deloop{\chop A}}{\deloop{\chop t}}{\deloop{\chop u}}
      \end{align*}
      and by induction $A[\bullet_\GD] = \deloop{\chop A}$,
      $t[\bullet_\GD]=\deloop{\chop t}$ and
      $u[\bullet_\GD] = \deloop{\chop u}$.
    \end{itemize}
  \item \emph{Induction on terms}:
    \begin{itemize}
    \item For a variable $\GD\vdash x:A$, by definition the mapping
      $x\mapsto \deloop{\chop x}$ is the only mapping for $x$ in
      $\red{\GD}$, hence $x[\red\GD] = \deloop{\chop{x}}$.
    \item For a term of the form
      $\GD\vdash\constr_{\GG,A}[\Gg]:A[\Gg]$, we have the following
      equations
      \begin{align*}
        \constr_{\GG,A}[\Gg][\red\GD] &= \constr_{\GG,A}[\Gg\circ\red\GD] &
                                                                            \deloop{\chop \constr_{\GG,A}[\Gg]} &=
                                                                                                                  \constr_{\GG,A}[\red\GG\circ\deloop{\chop \Gg}]
      \end{align*}
      and the induction case for substitutions then gives the
      equality.
    \end{itemize}
  \item \emph{Induction case for substitutions}:
    \begin{itemize}
    \item For the empty substitution $\GD\vdash\sub{}:\emptycontext$,
      since $\red\emptycontext = \sub{}$, we have the equalities
      \begin{align*}
        \sub{}\circ\red\GD &= \sub{} &
                                       \red\emptycontext\circ \deloop{\chop {\sub{}}} &= \sub{}
      \end{align*}
    \item For of the substitution
      $\GD\vdash\sub{\Gg,x\mapsto t} : (\GG,x:A)$, since the
      substitution $\red\GG$ has source $\deloop{\chop \GG}$ that do
      not use the variable $x$, we have
      $\red\GG\circ\sub{\deloop{\chop \Gg},x\mapsto\deloop{\chop t}} =
      \red\GG\circ\deloop{\chop \Gg}$. Moreover, if $x$ is of type
      $\Obj$ we have
      $\deloop{\chop x}[\sub{\deloop{\chop \Gg},x\mapsto\deloop{\chop
          t}}] = \new{} = \deloop{\chop t}$, and if $x$ is not of type
      $\Obj$, the expression
      $x[\sub{\deloop{\chop \Gg},x\mapsto\deloop{\chop
          t}}]=\deloop{\chop t}$. Using the two previously shown
      equalities to simplify the expressions yields
      \begin{align*}
        \sub{\Gg,x\mapsto t}\circ\red\GD &= \sub{\Gg\circ\red\GD,x\mapsto t[\red\GD]}&
                                                                                       \red{(\GG,x:A)}\circ\sub{\deloop{\chop \Gg},x\mapsto \deloop{\chop t}} &= \sub{\red\GG\circ \deloop{\chop{\Gg}},x\mapsto \deloop{\chop t}}
      \end{align*}
      By induction we have
      $\Gg\circ\red\GD = \red\GG\circ\deloop{\chop \Gg}$ and
      $t[\red\GD] = \deloop{\chop t}$ \qedhere
    \end{itemize}
  \end{induction}
\end{proof}

\begin{lemma}\label{lemma:red-correct}
  The following rule is admissible {\small
    \[
      \inferrule{\GG\vdash_{\CaTT}}{\deloop{\chop \GG}\vdash_{\CaTT}
        \red\GG:\GG}
    \]}
\end{lemma}
\begin{proof}
  This result is proved by induction on the context $\GG$.
  \begin{itemize}
  \item For the context $\emptycontext$, we have already proven
    that $\deloop{\chop \emptycontext}\vdash$. The
    rule~\regle{es} proves that
    $\deloop{\chop \emptycontext}\vdash \sub{}:\emptycontext$.
  \item For the context $(\GG,x:A)\vdash$, we have by
    induction $\deloop{\chop{\GG}}\vdash \red\GG : \GG$. By
    Proposition~\ref{prop:properties-tt}, this gives
    a derivation of
    $\deloop{\chop{(\GG,x:A)}}\vdash \red\GG:\GG$. Moreover, by
    Lemma~\ref{lemma:rsusp-correct-gsett}, we have that
    $\deloop{\chop{(\GG,x:A)}}\vdash x:\deloop{\chop A}$, and
    Lemma~\ref{lemma:syn-red} then shows that we have
    $\deloop{\chop{(\GG,x:A)}}\vdash x:A[\bullet_\GG]$. The rule~\regle{se} then gives a derivation of
    $\deloop{\chop{(\GG,x:A)}}\vdash\bullet_{(\GG,x:A)}:(\GG,x:A)$.\qedhere
  \end{itemize}
\end{proof}

\paragraph{Correctness of the reduced suspension.}
We are now equipped to generalize the reduced suspension as an
operation from the theory $\MCaTT$ to the theory $\CaTT$. We prove the
following correctness result showing that this operation is a well
defined translation between these two theories.
\begin{prop}\label{prop:rsusp-correct}
  The reduced suspension operation preserves derivability, the
  following rules are admissible {\small
    \[
      \inferrule{\GD\vdash_{\MCaTT}}{\deloop{\GD\vdash_{\CaTT}}}
      \qquad\qquad
      \inferrule{\GD\vdash_{\MCaTT}A}{\deloop{\GD\vdash_{\CaTT}}\deloop
        A} \qquad\qquad
      \inferrule{\GD\vdash_{\MCaTT}t:A}{\deloop{\GD\vdash_{\CaTT}}\deloop
        t:\deloop A} \qquad\qquad
      \inferrule{\GD\vdash_{\MCaTT}\Gg:\GG}{\deloop{\GD\vdash_{\CaTT}}\deloop\Gg
        : \deloop\GG}
    \]}
\end{prop}
\begin{proof}
  The proof of this result is essentially the same as the proof of
  Lemma~\ref{lemma:rsusp-correct-gsett}, but still needs some
  adaptations. For the sake of completeness and to avoid the reader to
  constantly refer to the previous proof, we give the complete proof
  here. We perform mutual induction on the derivation trees and keep
  all the cases in normal form.
  \begin{induction}
  \item \emph{Induction for contexts}:
    \begin{itemize}
    \item For the empty context $\emptycontext$, we have
      $\deloop{\emptycontext} = (\new {}:\Obj)$, and we can construct
      a derivation of $\deloop{\emptycontext}\vdash$ by applying
      successively the rules \regle{ec},\regle{$\Obj$-intro} and
      \regle{ce}.
    \item For the context $(\GG,x:\unit)\vdash$, we have
      $\deloop{(\GG,x:\unit)} = \deloop{\GG}$ and by induction
      $\deloop{\GG}\vdash$.
    \item For the context $(\GG,x:A)\vdash$ with $A\neq\unit$, we have
      $\GG\vdash A$. By induction this gives a derivation for
      $\deloop\GG\vdash \deloop A$. The rule~\regle{ce}, provides a
      derivation for $(\deloop\GG,x:\deloop A)\vdash$
    \end{itemize}
  \item \emph{Induction for types}:
    \begin{itemize}
    \item For the type $\GG\vdash\unit$, we have a derivation of
      $\GG\vdash$. By induction, we have a derivation of
      $\deloop\GG\vdash$. The rule~\regle{$\Obj$-intro} then applies
      to provide a derivation of $\deloop\GG\vdash\Obj$.
    \item For the type $\GG\vdash \Hom Atu$, we have derivations of
      $\GG\vdash A$, $\GG\vdash t:A$ and $\GG\vdash u:A$. By
      induction, those give derivations of
      $\deloop\GG\vdash\deloop A$,
      $\deloop\GG\vdash \deloop t : \deloop A$ and
      $\deloop \GG\vdash \deloop u : \deloop A$.  The
      rule~\regle{$\Hom{}{}{}$-intro} then provides a derivation of
      $\deloop\GG\vdash \Hom{\deloop A}{\deloop t}{\deloop u}$.
    \end{itemize}
  \item \emph{Induction for terms}:
    \begin{itemize}
    \item For the term $\GG\vdash \texttt{()}:\unit$, we have
      $\deloop{\texttt{()}} = \new{}$. By induction, we have
      $\deloop\GG\vdash$. Since moreover $(\new{}:\Obj)\in\deloop\GG$,
      the rule~\regle{var} gives a derivation of
      $\deloop\GG\vdash \new{} : \Obj$.
    \item For a variable $\GG\vdash x:A$, since we consider normal
      forms, $A\neq\unit$.  We have a derivation of $\GG\vdash$, and
      by induction it gives a derivation of
      $\deloop\GG\vdash$. Moreover, the condition $(x:A)\in\GG$ is
      satisfied, hence so is $(x:\deloop A)\in\deloop\GG$. The
      rule~\regle{var} then yields a derivation of
      $\deloop\GG\vdash x:\deloop A$.
    \item For a term of the form $\mconstr_{\GG,A}[\Gg]$, we have a
      derivation of $\GD\vdash \Gg:\chop \GG$. By induction, this
      gives a derivation for
      $\deloop \GD\vdash \deloop\Gg:\deloop{\chop \GG}$.  Moreover,
      Lemma~\ref{lemma:red-correct} ensures that we have
      $\deloop{\chop \GG}\vdash \red\GG : \GG$, hence we have a
      substitution $\deloop \GD\vdash \red\GG\circ\deloop\Gg :
      \GG$. By applying the rule~\regle{$\cohop$-intro}
      or~\regle{$\coh$-intro}, this provides a derivation for the
      judgment
      $\deloop\GD\vdash\constr_{\GG,A}[\red\GG\circ\deloop\Gg] :
      A[\red\GG\circ\deloop\Gg]$. The following equalities then prove
      that the type is the one we expect
      \begin{align*}
        A[\red\GG\circ \deloop\Gg] &= \deloop{\chop A}[\deloop \Gg] & \text{By Lemma~\ref{lemma:syn-red}} \\
                                   &= \deloop{((\chop A)[\Gg])} & \text{By Lemma~\ref{lemma:syn-rsusp}}
      \end{align*}
    \end{itemize}
  \item \emph{Induction for substitutions}:
    \begin{itemize}
    \item For the substitution $\GD\vdash\sub{}:\emptycontext$, we
      have a derivation of $\GD\vdash$. By induction, this gives a
      derivation of $\deloop\GD\vdash$. Moreover, we also have by
      definition that $(\new {}:\Obj)\in\deloop\GD$, and we have
      already constructed a derivation of $(\new{}:\Obj)\vdash$. This
      lets us construct a derivation for
      $\deloop\GD \vdash \deloop{\sub{}}:\deloop{\emptycontext}$ as
      follows {\small
        \[
          \inferrule*[Right=\regle{se}]{
            \inferrule*[Right=\regle{es}]{\deloop\GD\vdash}{\deloop{\GD}\vdash\sub{}:\emptycontext}\\
            (\new {}:\Obj)\vdash \\
            \inferrule*[Right=\regle{var}]{\deloop\GD\vdash \\
              (\new {}:\Obj)\in\deloop\GD}{\deloop\GD\vdash\new
              {}:\Obj}} {\deloop\GD\vdash\sub{\new {}\mapsto\new {}} :
            (\new {}:\Obj)}
        \]}
    \item For the substitution
      $\GD\vdash\sub{\Gg,x\mapsto t}:(\GG,x:\unit)$, we have the
      equalities
      \begin{align*}
        \deloop{(\GG,x:\unit)} &= \deloop \GG &
                                                \deloop{\sub{\Gg,x\mapsto t}} &= \deloop\Gg
      \end{align*}
      Since we have $\GD\vdash \Gg:\GG$, by induction we deduce
      $\deloop\GD\vdash\deloop\Gg:\deloop\GG$.
    \item For the substitution
      $\GD\vdash\sub{\Gg,x\mapsto t}:(\GG,x:A)$, we have $A \neq\unit$
      since we are in $\Glob$. Then we necessarily have derivations of
      $\GD\vdash \Gg:\GG$, $(\GG,x:A)\vdash$ and $\GD\vdash t:A[\Gg]$.
      By induction, those give derivations of the judgments
      $\deloop\GD\vdash\deloop\Gg:\deloop\GG$,
      $\deloop\GG,x:\deloop A\vdash$ and
      $\deloop{\GD}\vdash \deloop t:
      \deloop{A[\Gg]}$. Lemma~\ref{lemma:syn-rsusp} allows to rewrite
      the last of these judgments as
      $\deloop\GD\vdash \deloop t:\deloop A[\deloop \Gg]$. The
      rule~\regle{se} then provides a derivation of
      $\deloop\GD\vdash\deloop{\sub{\Gg,x\mapsto
          t}}:\deloop{(\GG,x:A)}$. \qedhere
    \end{itemize}
  \end{induction}
\end{proof}

\paragraph{Examples.}
We give a few examples of contexts in $\MCaTT$ along with their translations
in $\CaTT$ to build intuition on this translation:
\begin{align*}
  \GG &= (x:\Obj) & \deloop\GG &= (\new{}:\Obj, x:\Hom\Obj{\new{}}{\new{}})\\
  \GG &= (x:\Obj,f:\Hom\Obj xx, a:\unit) & \deloop\GG &= (\new{}:\Obj,x:\Hom\Obj{\new{}}{\new{}},f:\Hom{\Hom{}{\new{}}{\new{}}}xx)\\
  \GG &= (x:\Obj,y:\Obj;f:\Hom\Obj xy) & \deloop\GG &= (\new{}:\Obj,x:\Hom\Obj{\new{}}{\new{}},y:\Hom\Obj{\new{}}{\new{}},f:\Hom{\Hom{}{\new{}}{\new{}}}{x}{y})
\end{align*}
These examples help understand why it is important to require that the
translation is defined on normal forms. Indeed, in the second example, we have a
derivation $\GG\vdash a:\unit$ but the term is not in normal form. Putting it in
normal form before applying the reduced suspension yields to the term $\GG\vdash
\new{}:\unit$ which is indeed derivable. However, had we simply defined the
reduced suspension of any variable to be itself, without consideration of
whether it is in normal form, we would have gotten the judgment $\deloop
\GG\vdash a:\Obj$ which is not derivable.

\paragraph{Reduced suspension as a functor.}
Like for the desuspension, we can reformulate this result in a more
categorical flavor. However the result that we prove here is weaker.
\begin{coro}\label{coro:rsusp-functor}
  The reduced suspension operation defines a functor
  $\deloop{} : \Syn{\MCaTT} \to \Syn{\CaTT}$.
\end{coro}
\begin{proof}
  We have proved in Lemma~\ref{lemma:syn-rsusp} and
  Proposition~\ref{prop:desusp-correct} that this operation is well
  defined, and that it preserves the composition of substitutions, so
  it suffices to prove that it preserves the identity. We proceed by
  induction on the length of the context
  \begin{itemize}
  \item For the empty context $\emptycontext$, we have
    $\id{\emptycontext} = \sub{}$ along with
    $\deloop{\emptycontext} = (\new{}:\Obj)$ and
    $\deloop{\sub{}}=\sub{\new{}\mapsto \new{}}$, which is the
    identity of $\deloop{\emptycontext}$.
  \item For a context of the form $(\GG, x:\unit)$, we have
    $\id{(\GG,x:\unit)} = \sub{\id\GG,\new{}\mapsto\new{}}$. Hence
    $\deloop{\id{(\GG,x:\unit)}} = \deloop{\id{\GG}}$, and by
    induction $\deloop{\id\GG} = \id{\deloop\GG}$. We conclude using
    the fact that $\deloop{(\GG,x:\unit)} = \deloop\GG$.
  \item For a context of the form $(\GG,x:A)$ with $A\neq\unit$, we
    have $\deloop{\id{(\GG,x:A)}} = \sub{\deloop{\id\GG},x\mapsto x}$,
    and by induction, we have $\deloop{\id\GG} = \id{\deloop \GG}$,
    which lets us conclude that
    $\deloop{\id{(\GG,x:A)}} = \id{\deloop{(\GG,x:A)}}$. \qedhere
  \end{itemize}
\end{proof}
\noindent Note that in fact Lemma~\ref{lemma:syn-rsusp} and
Proposition~\ref{prop:desusp-correct} provide more structure that
this, they show that this functor is almost a morphism of category
with families. In fact it defines a morphism in $\Cat/\Fam$, which
respects the structures of categories with families of $\Syn{\MCaTT}$
and $\Syn{\CaTT}$. This functor however fails to be a morphism of
categories with families as it does not preserve the terminal object,
since the context $\emptycontext$ is sent onto $(\new{}:\Obj)$ which
is not terminal. This result however is complicated to formulate
cleanly in a categorical way.

\subsection{Interaction between desuspension and reduced suspension}
We have defined the theory $\MCaTT$ together with two translations, the
desuspension and the reduced suspension, that define functors between their
syntactic categories. We now study how these translations relate to each other
syntactically and translate this result categorically as a relation between the
two functors. In order to understand the interaction between the desuspension
and the reduced suspension, we first show the following result,
analogous to Lemma~\ref{lemma:unit-terminal} for the theory $\MCaTT$.
\begin{lemma}\label{lemma:contextplusone}
  The context $(x:\unit)$ is terminal in the category $\Syn{\MCaTT}$,
  and more generally the family of substitutions
  $\GG\vdash \sub{\id\GG,x\mapsto ()} : (\GG,x:\unit)$ defines a
  natural isomorphism.
\end{lemma}
\begin{proof}
  The proof of the initiality of the context $(x:\unit)$ is the exact
  same that the one of Lemma~\ref{lemma:unit-terminal}. The
  generalization of this statement is obtained by noticing that we
  have the following pullback square
  \[
    \begin{tikzcd}
      (\GG,x:\unit) \ar[r,"\sub{x,\mapsto x}"]\ar[d]\ar[dr,phantom,"\lrcorner" very near start] & (x:\unit)\ar[d] \\
      \GG \ar[r] & \emptycontext
    \end{tikzcd}
  \]
  Since the map $(x : \unit) \to \emptycontext$ is an isomorphism
  whose inverse is given by the substitution
  $\emptycontext \vdash \sub{x\mapsto ()} : (x:\unit)$, in this
  pullback implies that the left most vertical arrow is an
  isomorphism. Moreover, the structure of category with families
  implies that the inverse of this map is given by the substitution
  $\sub{\id\GG,x\mapsto ()}$. The family of morphisms defined this way
  is natural since the substitution is functorial, as it is defined by
  a universal property.
\end{proof}

\paragraph{Desuspension of the reduced suspension.}
We can now express one of the relations that link the desuspension and
the reduced suspension. We first express these relations on the type
theory before leveraging the expressive power of the categorical
semantics to provide a much more compact reformulation.
\begin{prop}\label{prop:chop-deloop}
  There exists a natural transformation
  $\eta : \id{\Syn{\MCaTT}} \to \chop{}\circ\deloop{}$ which is
  natural isomorphism.
\end{prop}
\begin{proof}
  We define a family of maps $\eta_\GG : \GG \to \chop{\deloop \GG}$
  for every context $\GG$ of the theory $\MCaTT$, and show that these
  maps are isomorphisms and that they are natural in $\GG$. We proceed
  by mutual induction on the theory $\MCaTT$ and prove
  \begin{itemize}
  \item For all context $\GG\vdash$, there is an isomorphism
    $\eta_\GG : \GG \to \chop{\deloop{\GG}}$.
  \item For all type $\GG\vdash A$, we have the equality
    $\GG\vdash A \equiv \chop{\deloop{\GG}}[\eta_\GG]$.
  \item For all term $\GG\vdash t:A$, we have
    $\GG\vdash t \equiv \chop{\deloop t}[\eta_\GG]:A$.
  \item For all substitution $\GD\vdash \Gg:\GG$, the following square
    commutes
    \[
      \begin{tikzcd}[row sep = small]
        \chop{\deloop\GD}\ar[r, "\chop{\deloop\Gg}"] \ar[-,d,"\eta_\GD\rotatebox{90}{\(\sim\)}"']& \chop{\deloop \GG} \ar[-,d,"\rotatebox{90}{\(\sim\)}\eta_\GG"]\\
        \GD\ar[r, "\Gg"''] & \GG
      \end{tikzcd}
    \]
  \item (\emph{auxiliary induction case}) for all context $\GG$ in the
    theory $\CaTT$, we have
    $\chop{\red{\GG}}\circ\eta_{\chop\GG} = \id{\chop\GG}$.
  \end{itemize}
  \begin{induction}
  \item \emph{Induction for contexts}:
    \begin{itemize}
    \item For the empty context $\emptycontext\vdash$, we have
      $\deloop{\emptycontext} = (\new {}:\Obj)$, and thus
      $\chop{\deloop{\emptycontext}} =
      (\new{}:\unit)$. Lemma~\ref{lemma:contextplusone} then provides
      the desired isomorphism
      $\eta_\emptycontext=\sub{\new{}\mapsto ()} : \emptycontext \to
      \chop{\deloop{\emptycontext}}$.
    \item For a context of the form $(\GG,x:\unit)\vdash$, we have
      $\deloop{(\GG,x:\unit)} = \deloop\GG$, hence
      $\chop{\deloop{(\GG,x:\unit)}} = \chop{\deloop{\GG}}$.  By
      Lemma~\ref{lemma:contextplusone}, the following weakening is an
      isomorphism $\pi : (\GG,x:\unit)\overset{\sim}\to \GG$, and the
      induction hypothesis on contexts provides the isomorphism
      $\eta_\GG : \GG \overset{\sim}\to \chop{\deloop{\Gg}}$. By
      composition, this provides the isomoprhism following (writing
      the weakening implicitly)
      $\eta_{(\GG,x:\unit)} = \eta_\GG : (\GG,x:\unit) \to
      \chop{\deloop{(\GG,x:\unit)}}$.
    \item For a context of the form $(\GG,x:A)\vdash$ with
      $A\neq\unit$, we have
      $\deloop{(\GG,x:A)} = (\deloop\GG,x:\deloop A)$. Since
      $\deloop A$ is necessarily distinct from $\Obj$, we have
      $\chop{\deloop {(\GG,x:A)}} = (\chop{\deloop
        \GG},x:\chop{\deloop A})$, and the induction case for contexts
      provides the isomoprhism
      $\eta_\GG : \GG\overset{\sim}\to\chop{\deloop\GG}$. Moreover,
      the induction case for types shows that
      $\GG\vdash A\equiv \chop{\deloop A}[\eta_\GG]$. Hence we define
      the isomoprhism
      $\eta_{(\GG,x:A)} = \sub{\eta_\GG,x\mapsto x} :
      (\GG,x:A)\overset\sim\to \chop{\deloop{(\GG,x:A)}}$, whose
      inverse is given by $\sub{\eta_\GG^{-1} , x\mapsto x}$.
    \end{itemize}
  \item \emph{Induction for types}:
    \begin{itemize}
    \item For the type $\GG\vdash \unit$, we have
      $\chop{\deloop\unit}[\eta_\GG] = \unit[\eta_\GG]$. Since we have
      also by definition $\unit[\eta_\GG] = \unit$, this shows that
      $\unit = \chop{\deloop\unit}[\eta_\GG]$.
    \item For the type $\Obj = \Hom{\unit}{()}{()}$, we have
      $\chop{\deloop \Obj}[\eta_\GG] =
      \Hom{\unit}{\new{}[\eta_\GG]}{\new{}[\eta_\GG]}$. Note that by
      Proposition~\ref{prop:rsusp-correct} along with the induction
      case for contexts, we have a derivation of
      $\GG\vdash \new{}[\eta_\GG]:\unit$. Hence the
      rule~\regle{$\eta_{\unit}$} applies and provides the equality
      $\GG\vdash \Obj\equiv \chop{\deloop\Obj}[\eta_\GG]$.
    \item For a type of the form $\GG\vdash \Hom Atu$ with
      $A\neq\unit$,
      $\chop{\deloop {\Hom Atu}} = \Hom{\chop{\deloop
          A}}{\chop{\deloop t}}{\chop{\deloop u}}$. The induction case
      for types then shows that
      $\GG\vdash A\equiv\chop{\deloop A}[\eta_\GG]$, and the induction
      case for terms shows
      $\GG\vdash t:A\equiv \chop{\deloop t}[\eta_\GG]$ similarly for
      $u$, which lets us conclude.
    \end{itemize}
  \item \emph{Induction for terms}:
    \begin{itemize}
    \item For the term $\GG\vdash ():\unit$, we have
      $\GG\vdash \chop{\deloop{()}}[\eta_\GG] : \unit$, and the rule
      \regle{$\eta_{\unit}$} gives the desired definitional equality.
    \item For a variable $\GG\vdash x:A$, we need to have $A\neq\unit$
      for the term to be in normal form, we have $\deloop x = x$ and
      thus $\chop{\deloop x} = x$. Moreover, since the pair $x:A$
      appears in $\GG$ with $A\neq\unit$, by definition of $\eta_\GG$
      (induction case for contexts), the association $x\mapsto x$
      appears in $\eta_\GG$, and hence
      $\chop{\deloop x}[\eta_\GG] = x$.
    \item For a term of the form $\mop_{\GG,A}[\Gg]$, we have
      $\deloop{\mop_{\GG,A}[\Gg]} =
      \cohop_{\GG,A}[\red\GG\circ\deloop\Gg]$, and thus we have the
      following equalities
      \begin{align*}
        (\chop{\deloop{\mop_{\GG,A}[\Gg]}})[\eta_\GD] &= \mop_{\GG,A}[\chop{(\red\GG\circ\deloop\Gg)}\circ \eta_\GD] \\
                                                      &= \mop_{\GG,A}[\chop{\red\GG}\circ(\chop{\deloop \Gg}\circ \eta_\GD)] & \text{by Lemma~\ref{lemma:syn-desusp}}\\
                                                      &\equiv \mop_{\GG,A}[(\chop{\red\GG}\circ \eta_{\chop \GG}) \circ \Gg] & \text{by the induction case for substitutions} \\
                                                      &\equiv \mop_{\GG,A}[\Gg] & \text{by the auxiliary induction case}
      \end{align*}
      Note that the that in the third step, the substitution
      $\eta_{\chop\GG}$ appears since we started with a definition
      $\GG$ whose target is $\chop\GG$.
    \item The case for a term of the form $\mcoh_{\GG,A}[\Gg]$ follows
      the exact same steps.
    \end{itemize}
  \item \emph{Induction for substitutions}:
    \begin{itemize}
    \item For the empty substitution $\GD\vdash \sub{}:\emptycontext$,
      by the induction case for contexts, we have the following square
      of maps
      \[
        \begin{tikzcd}
          \GD \ar[r,"\sub{}"]\ar[d,"\eta_\GD\rotatebox{90}{\(\sim\)}"'] & \emptycontext \ar[d,,"\rotatebox{90}{\(\sim\)}\eta_\emptycontext"]\\
          \chop{\deloop{\GD}} \ar[r,"\chop{\deloop{\sub{}}}"'] &
          (x:\unit)
        \end{tikzcd}
      \]
      Since by Lemma~\ref{lemma:contextplusone} both $\emptycontext$
      and $(x:\unit)$ are terminal objects, this square has to
      commute.
    \item For a substitution of the form
      $\GD\vdash \sub{\GG,x\mapsto t}:(\GG,x:\unit)$, after applying
      the induction case for contexts and computing the expressions,
      we are left with the following square.
      \[
        \begin{tikzcd}
          \GD \ar[r,"\sub{\Gg,x\mapsto t}"]\ar[d,"\eta_\GD\rotatebox{90}{\(\sim\)}"'] & (\GG,x:\unit) \ar[d,,"\rotatebox{90}{\(\sim\)}\eta_{(\GG,x:\unit)}"]\\
          \chop{\deloop{\GD}} \ar[r,"\chop{\deloop{\Gg}}"'] &
          \chop{\deloop\GG}
        \end{tikzcd}
      \]
      Note that by definition, the map
      $\eta_{(\GG,x:\unit)} = (\GG,x:\unit) \to \chop{\deloop \GG}$ is
      the map $\eta_\GG$ weakened with respect to the variable $x$,
      hence we have
      $\eta_{(\GG,x:\unit)} \circ \sub{\Gg,x\mapsto t} = \eta_\GG
      \circ \Gg$. By the induction case for substitutions, this
      expression is equal to $\chop{\deloop\Gg} \circ\eta_\GD$, which
      exactly provides the commutation of the above square.
    \item For a substitution of the form
      $\GD\vdash\sub{\Gg,x\mapsto t}:(\GG,x:A)$ with $A\neq\unit$, we
      have the following equalities
      \begin{align*}
        \sub{\chop{\deloop\Gg},x\mapsto \chop{\deloop t}}\circ \eta_\GD
        &= \sub{\chop{\deloop\Gg}\circ\eta_\GD, x\mapsto \chop{\deloop t}[\eta_\GD]}\\
        &= \sub{\eta_\GG\circ \Gg, x\mapsto t} &\text{By induction on terms and substitutions}\\
        &=\sub{\eta_\GG,x\mapsto x} \circ \sub{\Gg,x\mapsto t} &\text{Since $\eta_\GG$ is weakened with respect to $x$}
      \end{align*}
      which give exactly the commutation of desired naturality square.
    \end{itemize}
  \item\emph{Auxiliary induction case:} We prove by induction on
    contexts that $\chop{\red\GG}\circ\eta_{\chop\GG} = \id{\chop\GG}$
    \begin{itemize}
    \item For the context $\emptycontext$ in $\CaTT$, the map
      $\chop{\red\emptycontext}\circ\eta_{\chop\emptycontext}$ defines
      a map $\chop\emptycontext\to\chop\emptycontext$. By definition,
      $\chop\emptycontext$ is the terminal object in the category
      $\Syn{\MCaTT}$, the aforementioned map is thus necessarily the
      identity.
    \item For the context $(\GG,x:A)$ in $\CaTT$, we have the
      following equalities
      \begin{align*}
        \chop{\red{(\GG,x:A)}}\circ\eta_{\chop{(\GG,x:A)}}
        &= \sub{\chop{\red\GG},\chop{\deloop{\chop x}}} \circ \eta_{\chop\GG} \\
        &= \sub{\chop{\red\GG}\circ\eta_\GG, \chop{\deloop x}[\eta_{\chop\GG}]} &\text{Since $\chop x = x$}\\
        &= \sub{\id\GG,x\mapsto x} &\text{By induction on terms}
      \end{align*} \qedhere
    \end{itemize}
  \end{induction}
\end{proof}

\paragraph{A~natural transformation.}
In order to study the reduced suspension, we have introduced the family of
substitutions $\red\GG$ for all contexts $\GG$, and we have proved in
Lemma~\ref{lemma:red-correct}, that it defines a family of morphisms
$\red\GG:\deloop{\chop \GG}\to\GG$. Moreover, the equality $\Gg\circ\red\GD =
\red\GG\circ\deloop{\chop \Gg}$ that we proved in Lemma~\ref{lemma:syn-red} for
all substitution $\Gg$ can be expressed by the commutation of the following
diagram
\[
  \begin{tikzcd}
    \deloop{\chop \GD} \ar[r,"\red\GD"]\ar[d,"\deloop{\chop \Gg}"'] & \GD\ar[d,"\Gg"] \\
    \deloop{\chop \GG} \ar[r,"\red\GG"']& \GG
  \end{tikzcd}
\]
So we have in fact already proven the following proposition, which is simply a
categorical reformulation of our previous fact
\begin{prop}\label{prop:deloop-chop}
  The family of morphisms $\red{\GG} : \deloop{\chop \GG}\to \GG$ defines a
  natural transformation $\deloop{}\circ\chop{} \Rightarrow \id{\Syn\CaTT}$.
\end{prop}

\paragraph{Adjunction between desuspension and reduced suspension.}
Combining Propositions~\ref{prop:chop-deloop}
and~\ref{prop:deloop-chop}, we have in fact proved the following
categorical result about the functors induced by desuspension and the
reduced suspension.
\begin{thm}\label{thm:coreflective-adjunction}
  The functor $\deloop{}$ is left adjoint to $\chop{}$, the counit is
  given by the family $\red\GG$ and the unit is the natural family of
  isomorphisms $\eta_\GG$. This adjunction is thus coreflective.
\end{thm}
\begin{proof}
  We have already proved that these two natural transformations
  $\red\GG$ and $\eta_\GG$ are well defined, and that $\eta_\GG$ is an
  isomorphism. So it suffices to prove that they satisfy the zigzag
  identities. In fact we have also already proved one of these, as the
  \emph{auxiliary induction case} while proving
  Proposition~\ref{prop:chop-deloop}. So we are left to check the
  other identity, namely that for all context $\GG$ in the theory
  $\MCaTT$, we have
  $\red{\deloop\GG}\circ \deloop{\eta_\GG} = \deloop\GG$.  We prove
  this statement by induction on contexts.
  \begin{itemize}
  \item For the empty context $\emptycontext$, since $\new{}$ is a
    variable, we have $\deloop{\chop{\new{}}} = \new{}$, and hence
    \[
      \red{\deloop{\emptycontext}}\circ\deloop{\eta_\emptycontext} =
      \red{(\new{},\Obj)}\circ\deloop{\sub{\new{}\mapsto ()}} =
      \sub{\new{}\mapsto
        \deloop{\chop{\new{}}}}\circ\sub{\new{}\mapsto\new{}} =
      \sub{\new{}\mapsto\new{}} = \id{\deloop{\emptycontext}}
    \]
  \item For the context $(\GG,x:\unit)$, we have by induction
    \[
      \red{\deloop{\GG,x:\unit}}\circ\deloop{\eta_{(\GG,x:\unit)}} =
      \red{\deloop{\GG}}\circ\deloop{\eta_{\GG}} = \id{\deloop{\GG}} =
      \id{\deloop{\GG,x:\unit}}
    \]
  \item For the context $(\GG,x:A)$ with $A\neq\unit$, we have
    \begin{align*}
      \red{\deloop{\GG,x:A}}\circ\deloop{\eta_{(\GG,x:A)}}
      &= \red{(\deloop{\GG},x:\deloop{A})}\circ\sub{\deloop{\eta_{\GG}},x\mapsto x} \\
      &=\sub{\red{\deloop\GG},x\mapsto \deloop{\chop x}} \circ \sub{\deloop{\eta_{\GG}},x\mapsto \deloop x}\\
      & = \sub{\red{\deloop\GG} \circ \deloop{\eta_\GG}, x\mapsto \deloop{\chop x} [\sub{\eta_\GG,x\mapsto \deloop x}]}
    \end{align*}
    Using the induction hypothesis, and the fact that since $x$ is a
    variable, we have $x = \chop x = \deloop{\chop x}$, it follows
    that
    $\red{\deloop{(\GG,x:A)}}\circ\deloop{\eta_{(\GG,x:A)}} =
    \id{(\deloop\GG,x:\deloop A)}$.
  \end{itemize}
\end{proof}

\noindent This adjunction is to be understood as an analogue in the
world of categories of the topological adjunction between the reduced
suspension and the loop space. In our terminology, the desuspension
corresponds to the loop space, and the reduced suspension corresponds
to the reduced suspension.

\begin{rem}\label{rem:coreflective-adjunction}
  The coreflective adjunction exhibits $\Syn{\MCaTT}$ as isomorphic to
  a coreflective subcategory of $\Syn{\CaTT}$. Moreover the essential
  image of $\deloop{}$ is exactly the category
  $\Syn{\CaTT,\bullet}$. So a way to understand our type theory
  $\MCaTT$ is that it achieves a structure of category with families
  on a category which is equivalent to $\Syn{\CaTT,\bullet}$, in such
  a way that that this structure coincide with the structure on
  $\CaTT$ under the equivalence between $\Syn{\MCaTT}$ and
  $\Syn{\CaTT,\bullet}$.
\end{rem}

\section{Models of the type theory $\MCaTT$.}
Thanks to the desuspension and to the reduced suspension functors that
we have defined establish a strong relation between the models of the
theory $\CaTT$ and those of the theory $\MCaTT$.

\subsection{Action of reduced suspention and desuspension on the models}
\paragraph{Desuspension acting on the models of $\MCaTT$.}
Consider a model $F : \MCaTT\to\Set$ and define the functor
$\chopstar{F} : \CaTT\to\Set$ by precomposing with the functor
$\chop{}$, so we have $\chopstar{F}(\GG) = F(\chop \GG)$ and
$\chopstar{F} (\Gg) = F(\chop \Gg)$. By
Corollary~\ref{coro:desusp-cwf-mor}, $\chop{}$ is a morphism of
categories with families, and those preserve display maps, the
terminal object and pullbacks along display maps. Since $F$ is a
model, it also preserves the terminal and the pullback along display
maps, hence so does the composite $\chopstar F$. Thus $\chopstar F$ is
a model of the theory $\CaTT$, and the desuspension induces a functor
$\chopstar{}:\Mod{\Syn{\MCaTT}}\to \Mod{\Syn{\CaTT}}$.  Moreover we
have $\chopstar{F}(x:\Obj) = F(x:\unit)$ and by
Lemma~\ref{lemma:contextplusone}, the context $(x:\unit)$ is terminal
and by definition, $F$ preserves the terminal object, it follows that
$\chopstar{F}(x:\Obj)$ is terminal in $\Set$, hence $\chopstar{F}$ is
an object of $\OneObjCat$. This shows that the desuspension induces a
functor$\chopstar{}:\Mod{\MCaTT} \to \OneObjCat$.

\paragraph{Reduced suspension acting on the models of $\CaTT$.}
We define a similar construction for the reduced suspension. However,
since the reduced suspension does not define an image for every term
and does not preserve the terminal object, the construction is
slightly more involved. We first consider start by showing the
following result
\begin{lemma}\label{lemma:deloopstar-preserves-dm}
  Given a model $F:\CaTT\to\Set$ of the type theory $\CaTT$, the
  functor $\deloopstar{F}$ preserves the pullbacks along display maps
  of $\Syn{\MCaTT}$.
\end{lemma}
\begin{proof}
  We consider a pullback along a display map, which is of the form
  \[
    \begin{tikzcd}
      (\GD,x:A[\Gg])\ar[dr,phantom,"\lrcorner" very near start] \ar[r]\ar[d,"\pi"'] & (\GG,x:A)\ar[d,"\pi"] \\
      \GD \ar[r,"\Gg"'] & \GG
    \end{tikzcd}
  \]
  We first consider the case $A=\unit$. In this case, the image by
  $\deloopstar{F}$ of the above square writes as
  \[
    \begin{tikzcd}
      \deloopstar{F}(\GD)\ar[dr,phantom,"\lrcorner" very near start] \ar[r,"\deloopstar{F}(\Gg)"]\ar[d,"\id{}"'] & \deloopstar{F}(\GG)\ar[d,"\id{}"] \\
      \deloopstar{F}(\GD) \ar[r,"\deloopstar{F}(\Gg)"'] & \deloopstar{F}(\GG)
    \end{tikzcd}
  \] which is a pullback. We now consider the case $A\neq\unit$, then
  the equality $\deloop{(A[\Gg])} = \deloop A[\deloop \Gg]$ given by
  Lemma~\ref{lemma:syn-rsusp} along with
  Lemma~\ref{lemma:pullback-display-map} shows that we have the
  following pullback square in $\Syn{\CaTT}$
  \[
    \begin{tikzcd}
      \deloop{(\GD,x:A[\Gg])}\ar[dr,phantom,"\lrcorner" very near start] \ar[r]\ar[d,"\pi"'] & \deloop{(\GG,x:A)}\ar[d,"\pi"] \\
      \deloop{\GD} \ar[r,"\deloop\Gg"'] & \deloop{\GG}
    \end{tikzcd}
  \]
  Since $F$ preserves pullbacks along display maps, this shows that
  the image of the initial square by $\deloopstar{F}$ is a pullback
  \[
    \begin{tikzcd}
      \deloopstar{F}(\GD,x:A[\Gg])\ar[dr,phantom,"\lrcorner" very near start] \ar[r]\ar[d,"\pi"'] & \deloopstar{F}(\GG,x:A)\ar[d,"\pi"] \\
      \deloopstar{F}(\GD) \ar[r,"\Gg"'] & \deloopstar{F}(\GG)
    \end{tikzcd}
  \] \qedhere
\end{proof}
This shows that $\deloopstar{F}$ is a model exactly when it preserves
the initial object. Since we have by definition
$\deloopstar F (\emptycontext) = F(\new{}:\Obj)$, this condition
translates to $F (\new{},\emptycontext)$ being a singleton.  Hence
$\deloopstar F$ is a model if and only if $F$ is an object of
$\OneObjCat$.  This shows that $\deloopstar{}$ defines a functor
$ \deloopstar{}:\OneObjCat \to \Mod{\Syn{\MCaTT}} $.

\subsection{Models of the category $\Syn{\MCaTT}$}
Using the desuspension and the reduced suspension, we have defined a
pair of functors $\chopstar{}$ and $\deloopstar{}$ between the
categories $\Mod{\Syn{\MCaTT}}$ and $\OneObjCat$.
\begin{thm}\label{thm:models-mcatt}
  The functors $\chopstar{}$ and $\deloopstar{}$ define an equivalence
  of categories between $\Mod{\Syn{\MCaTT}}$ and $\OneObjCat$
\end{thm}
\begin{proof}
  First, Proposition~\ref{prop:chop-deloop} implies
  $\deloopstar{}\circ\chopstar{} = \pa{\chop{}\circ\deloop{}}^* \simeq
  \id{\Syn{\MCaTT}}$ and Proposition~\ref{prop:deloop-chop} shows that
  there is a natural transformation, obtained by whiskering
  $\chopstar{}\circ\deloopstar{} = \pa{\deloop{}\circ\chop{}}^*
  \Rightarrow \id{\OneObjCat}$ So it suffices to show that this
  natural transformation is a natural isomorphism, that is for any
  $F\in\OneObjCat$ and all context $\GG$ in $\CaTT$, the map
  $F(\bullet_\GG) : F(\deloop{\chop{\GG}}) \to F(\GG)$ is an
  isomorphism. We prove this property by induction on the context
  $\GG$.
  \begin{itemize}
  \item For the empty context $\emptycontext$, we necessarily have
    that $F(\emptycontext) = \set{\bullet}$, and since
    $\deloop{\chop{\emptycontext}} = D^0$ and $F\in\OneObjCat$, we
    also have that $F(\deloop{\chop{\emptycontext}}) =
    \set{\bullet}$. Hence $F(\bullet_\emptycontext)$ is the unique map
    between the singleton to itself, which is an isomorphism.
  \item For a context of the form $(\GG,x:\Obj)$, it is obtained as a
    (trivial) pullback, and the map $\bullet_{(\GG,x:\Obj)}$ is
    obtained by universal property of the pullback, as follows
    
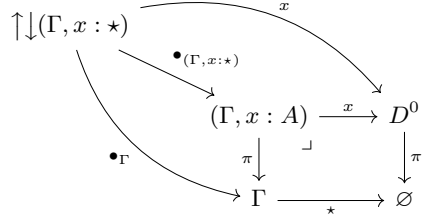
\[
      \begin{tikzcd}
        \deloop{\chop{(\GG,x:\Obj)}}\ar[ddr, "\bullet_\GG"', bend right]\ar[rrd, bend left, "x"]\ar[dr,"\bullet_{(\GG,x:\Obj)}"] & & \\
        & (\GG,x:A) \ar[d,"\pi"']\ar[r,"x"]\ar[dr,phantom,"\lrcorner" very near start] & D^0\ar[d,"\pi"] \\
        & \GG\ar[r,"\Obj"'] & \emptycontext
      \end{tikzcd}
    \]
    Taking the image by $F$ on this pullback yields another pullback
    in $\Set$ (since $F$ is a model of $\CaTT$), as follows
    
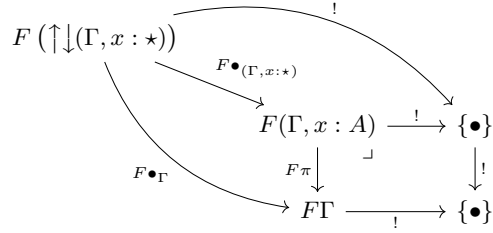
\[
      \begin{tikzcd}
        F \pa{\deloop{\chop{(\GG,x:\Obj)}}}\ar[ddr, "F\bullet_\GG"', bend right]\ar[rrd, bend left, "!"]\ar[dr,"F\bullet_{(\GG,x:\Obj)}"] & & \\
        & F(\GG,x:A) \ar[d,"F\pi"']\ar[r,"!"]\ar[dr,phantom,"\lrcorner" very near start] & \set{\bullet}\ar[d,"!"] \\
        & F\GG\ar[r,"!"'] & \set{\bullet}
      \end{tikzcd}
    \]
    Since the square is a pullback, $F\pi$ is an isomorphism, and
    since by induction $F\bullet_\GG$ is also an isomorphism,
    necessarily $F\bullet_{\GG,x:\Obj}$ is an isomorphism.
  \item For a context of the form $(\GG,x:A)$ where $A$ is a type
    distinct from $\Obj$, the context is obtained as a pullback, the
    context $\deloop{\chop(\GG,x:A)}$ is also a pullback and the
    substitution $\bullet_{(\GG,x:A)}$ is obtained by universal
    property as described in the following diagram which has the shape
    of a cube whose faces are commutative and whose front and back
    face are pullbacks
    
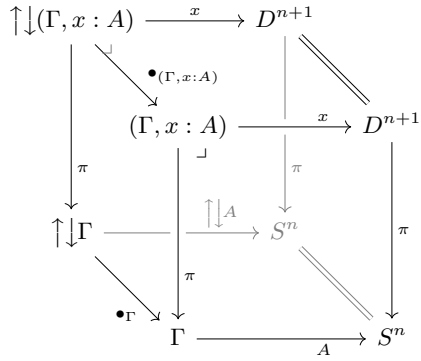
\[
      \begin{tikzcd}[row sep={40,between origins}, column
        sep={40,between origins}]
        \deloop{\chop{(\GG,x:A)}}\ar[rr,"x"]\ar[dd,"\pi"near
        end]\ar[dr,"\bullet_{(\GG,x:A)}"near end] \ar[ddrr, start anchor =
        {[xshift = -2.5em, yshift = .8em]south east},phantom,"\lrcorner" very
        near start, gray]
        & & D^{n+1}\ar[dr, equal]\ar[dd,"\pi" near end, gray] & \\
        & (\GG,x:A)\ar[rr,crossing over,"x", near end]\ar[ddrr, start anchor =
        {[xshift = -2.5em, yshift = .8em]south
          east},phantom,"\lrcorner" very near start] & & D^{n+1}\ar[dd,"\pi"]\\
        \deloop{\chop{\GG}} \ar[rr,"\deloop{\chop A}"near
        end, gray]\ar[dr,"\bullet_\GG"' near end] & & {\color{gray}S^n} \ar[dr,equal, gray] & \\
        & \GG \ar[rr,"A"'near end]\ar[from=uu, crossing over, "\pi" near end] &
        & S^n
      \end{tikzcd}
    \]
    Taking the image by $F$ of this diagram yields another cube whose
    faces are all commutative square, and whose front and back square
    are again pullback squares, since as a model, $F$ preserves the
    pullbacks along the display maps (in the following figure, we have
    left implicit most of the arrows, they are simply the image by $F$
    of the ones of the previous figure).
    \[
      \begin{tikzcd}[row sep={40,between origins}, column
        sep={40,between origins}]
        F(\deloop{\chop{(\GG,x:A)}})\ar[rr,"x"]\ar[dd,]\ar[dr,"F\bullet_{(\GG,x:A)}"near
        end] \ar[ddrr, start anchor = {[xshift = -2.5em, yshift = .8em]south
          east},phantom,"\lrcorner" very near start, gray]
        & & FD^{n+1}\ar[dr, equal]\ar[dd,gray] & \\
        & F(\GG,x:A)\ar[rr,crossing over]\ar[ddrr, start anchor = {[xshift =
          -2.5em, yshift = .8em]south
          east},phantom,"\lrcorner" very near start] & & FD^{n+1}\ar[dd]\\
        F(\deloop{\chop{\GG}}) \ar[rr, gray]\ar[dr,"F\bullet_\GG"'near end,
        "\mathclap{\sim}"sloped] & & {\color{gray}FS^n} \ar[dr,equal, gray] & \\
        & F\GG \ar[rr,"FA"'near end]\ar[from=uu, crossing over] & & FS^n
      \end{tikzcd}
    \]
    By induction, the map $F(\bullet_\GG)$ is an isomorphism, making
    the span defining $F(\GG,x:A)$ and the span defining
    $F(\deloop{\chop{(\GG,x:A)}})$ isomorphic. This proves that the
    map $F(\bullet_{(\GG,x:A)})$ is also an isomorphism, by uniqueness
    of the pullback up to isomorphism.
  \end{itemize}
\end{proof}

\paragraph{Reflective localization.}
We now give a reformulation of the construction we have
presented. First note that the opposite of a category with families
$C$ embeds inside its category of models via a Yoneda embedding
\[
  \begin{array}{rcl}
    \op{C} & \hookrightarrow & \Mod{C}\\
    \GG & \mapsto & C(\GG,\_)
  \end{array}
\]
Indeed, the functor $C(\GG,\_)$ preserves the pullbacks along display
maps (and in fact all limits) by continuity of the Hom-functor. This
lets us see the objects $\emptycontext$ and $D^0$ as particular
objects of $\Mod{\Syn{\CaTT}}$, with $\emptycontext$ being the initial
object. We then consider the unique map $s:\emptycontext \to D^0$ in
the category $\Mod{\Syn{\CaTT}}$. Then the category $\OneObjCat$ is
the category of all the $s$-local objects, \ie all objects $F$ such
that the map
\[
  \Mod{\Syn{\CaTT}}(s,F) : \Mod{\Syn{\CaTT}}(\emptycontext,F) \to
  \Mod{\Syn{\CaTT}}(D^0,F)
\] is a natural isomorphism. Indeed, since $\emptycontext$ is an
initial objects, $s$-local objects are exactly those such that
$\Mod{\Syn{\CaTT}}(D^0,F)$ is a singleton, which reformulates as
$F(D^0)$ being a singleton by the Yoneda lemma. This exhibits
$\OneObjCat$ as a reflective localization of the category
$\Mod{\Syn{\CaTT}}$ as $s$.  Theorem~\ref{thm:models-mcatt} then shows
that $\Mod{\Syn{\MCaTT}}$ is equivalent to $\OneObjCat$, and thus is
also a reflective localization of $\Mod{\Syn\CaTT}$ at $s$. This
localization lifts the coreflective adjunction between $\Syn{\CaTT}$
and $\Syn{\MCaTT}$ realized by the desuspension and the reduced
suspension, that we have stated in
Theorem~\ref{thm:coreflective-adjunction}.

\subsection{Interpretation}
As we have proved independently~\cite{benjamin2021globular}, the
models of the theory $\CaTT$ are equivalent to the
Grothendieck-Maltsiniotis definition of weak $\omega$-categories with
the Batanin-Leinster coherator. In light of this fact,
Theorem~\ref{thm:models-mcatt} can be reformulated as stating that the
models of $\MCaTT$ are the weak $\omega$-categories (in the sense of
Grothendieck-Malstiniotis with the Batanin-Leinster coherator) with a
single object. It is expected form higher category theory that such a
result holds~\cite{baezlecture}, however it is still surprising
that this definition makes sense at all, and we believe that the
construction that we have presented only makes sense for $1$-monoidal
category as is. To explain why, we assume that the syntactic category
$\Syn\CaTT$ is the opposite of the $\omega$-category freely generated
on finite computads, for an appropriate notion of computad, and
simiraly, that $\Syn{\MCaTT}$ is the opposite of the monoidal
$\omega$-categories freely generated by finite computads, for an
appropriate notion of computads for monoidal
$\omega$-categories. There is no reason a priori that a freely
generated monoidal $\omega$-category on a finite computad should also
be freely generated on a finite computad as a plain weak
$\omega$-category. We conjecture that it would not be the case for
$2$-monoidal categories: A freely generated monoidal weak
$\omega$-category has the monoidal unit $e$ as well as all the
iterated monoidal products of $e$. This is too much data to fit into a
context of the theory $\MCaTT$ (\ie a finite computad for monoidal
weak $\omega$-categories). Hence we expect that there is no freely
generated $2$-monoidal category that is also freely generated by a
finite computad as a monoidal $\omega$-category. However, considering
that our interpretation of the syntactic category is correct, we have
actually proved that every free monoidal $\omega$-category on a finite
computad is also free on a finite computad as a plain
$\omega$-category.

We also speculate that weak $\omega$-categories should be compared
with an appropriate notion of weak equivalence, that can be described
by a structure akin to a model structure, and similarly for the
monoidal weak $\omega$-categories. We have presented an equivalence
between the categories with strict functors, but we expect this
equivalence to lift and provide an equivalence between weak categories
with weak functors. We conjecture that the relevant property in this
weakened world is not having a single object (which is not expected to
be invariant under weak equivalences), but having a $0$-connected
groupoidal core. Investigation in this direction to straighten and
prove those claims are left for further work.


\section{Conclusion}
We have developed the type theory $\MCaTT$ whose models are monoidal
weak $\omega$\nobreakdash-ca\-te\-go\-ries. To this end, we rely on an existing
type theory describing weak $\omega$-categories and index the rule of our theory
with the rule of the existing theory. We were then able to prove correctness
results, using syntactic translations between the two theories and lifting the
results as a correspondence between their models.

\paragraph{A forgetful functor.}
There is another translation from $\Syn{\CaTT}$ to
$\Syn{\MCaTT}$ that we have not presented here. Given a ps-context $\GG$, we can
produce its \emph{suspension} $\GS \GG$, obtained by formally adding two objects
and lifting all the dimensions by $1$, in such a way that an object in $\GG$
becomes a $1$-cell between the two new objects in $\GS\GG$. We can then leverage
this operation to define a translation, which gives a morphism of categories
with families $u:\Syn{\CaTT}\to\Syn{\MCaTT}$.
The induced functor on the categories of models
\[
  u^* : \Mod{\Syn{\MCaTT}} \to \Mod{\Syn{\CaTT}}
\]
is the forgetful functor, which, given a monoidal weak
$\omega$-category, forgets the monoidal product and gives the
underlying weak $\omega$-category. Importantly, in this case there is
no shift of dimension; the objects of the $\omega$-category are the
objects of the monoidal $\omega$-category, but without the ability to
be composed. The existence of this functor is
closely related to the well-definedness of an operation of
\emph{suspension} in the theory $\CaTT$\cite{benjamin2019suspension}.

\paragraph{Alternate presentations of $\MCaTT$.}
We have worked out different presentations for the theory $\MCaTT$, all of them
being centered around the idea of enforcing the constraint of having a unique
object of dimension $-1$. We give in~\cite{these} two other presentations. The
first one is very similar to the one that we have given here, but the difference
is that we do not introduce a type $\unit$ in $\MCaTT$. The price to pay is that
the desuspension becomes a partial operation, since there is no type to send the
type $\Obj$ to, but this gets compensated by the fact that the theory can be
expressed without definitional equalities and thus there is no need to carefully
work with terms in normal forms. Both these presentation are relatively close,
and none of them really surpasses the other. The second presentation that we
give in~\cite{these} encodes all the properties with lists of contexts. This
presentation is more involved and significantly harder to study, however it
gives a syntax that is a bit more concise. It also provides a framework which is
more independent of $\CaTT$, and for which the conditions for constructing the
derivation trees are more straightforwardly verified. For these reasons, we
believe that the latter presentation is to be preferred for a potential future
implementation of the theory $\MCaTT$.

\paragraph{Monoidal closed higher categories.} It would be valuable to
connect the result we have presented with recent work of Finster for
integrating the theory of monoidal higher categories in the tool
Opetopic~\cite{opetopic}, although it necessitates to establish a connection
between globular shapes and opetopic shapes.

\bibliographystyle{plain}
\bibliography{article}

\newpage
\appendix
\section{Summary of the type theories}
\subsection{The type theory $\Glob$}\label{sec:app-glob}
\paragraph{Syntax.}
\begin{induction}
\item \emph{Contexts}: lists $(x_0:A_0,\ldots,x_n:A_n)$ with $x_i$ variables and
  $A_i$ types
\item \emph{Types}: either $\Obj$ or of the form $\Hom Atu$ with $A$ type and
  $t$ and $u$ terms
\item \emph{Terms}: variables
\item \emph{Substitutions}: lists $\sub{x_0\mapsto t_0,\ldots,x_n\mapsto t_n}$
  with $x_i$ variables and $t_i$ terms
\end{induction}

\paragraph{Inference rules.}
\[
  \begin{array}{c@{\qquad\qquad}c}
    \multicolumn{2}{l}{\text{\emph{For contexts}:}}\\
    \inferrule{\null}{\emptycontext\vdash}{\regle{ec}}  & \inferrule{\GG\vdash
                                                          A}{\GG,x:A\vdash}{\regle{ce}}
                                                          \quad \text{Where
                                                          $x\notin\Var\GG$}\\
    \multicolumn{2}{l}{\text{\emph{For types}:}}\\
    \inferrule{\GG\vdash}{\GG\vdash\Obj}{\regle{$\Obj$-intro}} & \inferrule{\GG\vdash t:A~\\ \GG\vdash u:A}{\GG\vdash\Hom Atu}{\regle{$\Hom{}{}{}$-intro}}\\
    \multicolumn{2}{l}{\text{\emph{For terms}:}}\\
    \inferrule{\Gamma\vdash \\ (x : A) \in \GG}{\Gamma\vdash x:A}{\regle{var}} \\
    \multicolumn{2}{l}{\text{\emph{For substitutions}:}}\\
    \inferrule{\GD\vdash}{\GD\vdash \sub{}:\emptycontext}{\regle{es}} & \inferrule{\GD\vdash \Gg: \GG \\ \GG,x:A\vdash~\\ \GD\vdash t:A[\Gg]}{\GD\vdash\sub{\Gg,x\mapsto t}:(\GG,x:A)}{\regle{se}}
  \end{array}
\]

\paragraph{Semantics.}
\begin{align*}
  \Syn{\Glob} &= \op{\FinGSet}\\
  \Mod{\Syn{\Glob}} &= \GSet
\end{align*}

\subsection{The type theory $\CaTT$}\label{sec:app-catt}
\paragraph{Syntax.}
\begin{induction}
\item \emph{Contexts}: lists $(x_0:A_0,\ldots,x_n:A_n)$ with $x_i$ variables and
  $A_i$ types
\item \emph{Types}: either $\Obj$ or of the form $\Hom Atu$ with $A$ type and
  $t$ and $u$ terms
\item \emph{Terms}: either variables or of the form $\cohop_{\GG,A}[\Gg]$ or $\coh_{\GG,A}[\Gg]$ with
  $\GG$ a ps-context, $A$ a type and $\Gg$ a substitution
\item \emph{Substitutions}: lists $\sub{x_0\mapsto t_0,\ldots,x_n\mapsto t_n}$
  with $x_i$ variables and $t_i$ terms
\end{induction}

\paragraph{Rules for ps-contexts.}
\[
  \begin{array}{c@{\qquad\qquad}c}
    \inferrule{\null}{(x : \Obj)\vdashps x : \Obj}{\regle{pss}}
    &
      \inferrule{\Gamma\vdashps x:A}{\Gamma,y:A,f\Hom Axy\vdashps f : \Hom Axy}{\regle{pse}}
    \\
    \inferrule{\Gamma\vdashps f : \Hom Axy}{\Gamma\vdashps y:A}{\regle{psd}}
    &
      \inferrule{\Gamma\vdashps x : \Obj}{\Gamma\vdashps}{\regle{ps}}
  \end{array}
\]

\paragraph{Source and target of a ps-context.}
\begin{align*}
  \partial^-_i(x:\Obj) &= (x:\Obj) &
                                     \partial^-_i(\GG,y:A,f:\Hom{}xy) &= \left\{
                                                                        \begin{array}{l@{\quad}l}
                                                                          \partial^-_i\GG & \text{if $\dim A~\geq i-1$} \\
                                                                          (\partial^-_i\GG,y:A,f:\Hom{}xy) & \text{otherwise}
                                                                        \end{array} \right. \\[2em]
  \partial_i^+(x:\Obj) &= (x:\Obj) &
                                     \partial^+_i(\GG,y:A,f:\Hom{}xy) &= \left\{
                                                                        \begin{array}{l@{\quad}l}
                                                                          \partial^+_i\GG & \text{if $\dim A~\geq i$} \\
                                                                          \mathrm{drop}(\partial^+_i\GG),y:A~& \text{if $\dim A~= i-1$}\\
                                                                          \partial^+_i\GG,y:A,f:\Hom{}xy & \text{otherwise}
                                                                        \end{array} \right. \\[2em]
  \src\GG &= \partial_{\dim \GG-1}^- \GG &  \tgt\GG &= \partial_{\dim\GG-1}^+\GG
\end{align*}
where $\mathrm{drop}(\GG)$ is the context $\GG$ with its last variable removed.

\paragraph{Side conditions.}
\[
  \inferrule{\GG\vdashps \\ \GG \vdash t:A \\ \GG\vdash
    u:A}{\GG\vdashop \Hom Atu}{\text{ when} \left\{
      \begin{array}{l}
        \Var t\cup\Var A = \Var{\src\GG} \\
        \Var u\cup\Var A = \Var{\tgt\GG}
      \end{array}
    \right.}
\]
\[
  \inferrule{\GG\vdashps \\ \GG \vdash t:A \\ \GG\vdash
    u:A}{\GG\vdasheq \Hom Atu}{\text{ when} \left\{
      \begin{array}{l}
        \Var t\cup\Var A = \Var{\GG} \\
        \Var u\cup\Var A = \Var{\GG}
      \end{array}
    \right.}
  \]

\paragraph{Inference rules.}
\[
  \begin{array}{c@{\qquad\qquad}c}
    \multicolumn{2}{l}{\text{\emph{For contexts}}:} \\
    \inferrule{\null}{\emptycontext\vdash}{\regle{ec}}  & \inferrule{\GG\vdash A}{\GG,x:A\vdash}{\regle{ce}} \quad \text{Where $x\notin\Var\GG$}\\
    \multicolumn{2}{l}{\text{\emph{For types}}:} \\
    \inferrule{\GG\vdash}{\GG\vdash\Obj}\regle{$\Obj$-intro} & \inferrule{\GG\vdash t:A~\\ \GG\vdash u:A}{\GG\vdash\Hom Atu}{\regle{$\Hom{}{}{}$-intro}}\\
    \multicolumn{2}{l}{\text{\emph{For terms}}:} \\
    \inferrule{\Gamma\vdash \\ (x : A) \in \GG}{\Gamma\vdash x:A}{\regle{var}} & \inferrule{\GG\vdashps \\ \GG\vdashop A \\ \GD\vdash\Gg:\GG}{\GD\vdash\cohop_{\GG,A}[\Gg]:A[\Gg]}{\regle{op}}\\
                                                        & \inferrule{\GG\vdashps \\ \GG\vdasheq A \\ \GD\vdash\Gg:\GG}{\GD\vdash\coh_{\GG,A}[\Gg]:A[\Gg]}{\regle{coh}}\\
    \multicolumn{2}{l}{\text{\emph{For substitutions}}:} \\
    \inferrule{\GD\vdash}{\GD\vdash \sub{}:\emptycontext}{\regle{es}} & \inferrule{\GD\vdash \Gg: \GG \\ \GG,x:A\vdash~\\ \GD\vdash t:A[\Gg]}{\GD\vdash\sub{\Gg,x\mapsto t}:(\GG,x:A)}{\regle{se}}
  \end{array}
\]

\paragraph{Semantics.}
\begin{align*}
  \Syn{\CaTT} & \text{ : oppoiste of finite computads for weak  $\omega$-categories} \\
  \Mod{\Syn{\CaTT}} & \text{ : equivalent to the category of weak $\omega$-categories}
\end{align*}

\subsection{The theory $\Glob_{\unit}$}\label{app:glob-unit}
\begin{induction}
\item \emph{Contexts}: lists $(x_0:A_0,\ldots,x_n:A_n)$ with $x_i$ variables and
  $A_i$ types
\item \emph{Types}: either $\unit$ or of the form $\Hom Atu$ with $A$ type and
  $t$ and $u$ terms
\item \emph{Terms}: variables
\item \emph{Substitutions}: lists $\sub{x_0\mapsto t_0,\ldots,x_n\mapsto t_n}$
  with $x_i$ variables and $t_i$ terms
\end{induction}

\paragraph{Inference rules.}
\[
  \begin{array}{c@{\qquad\qquad}c}
    \multicolumn{2}{l}{\text{\emph{For contexts}:}}\\
    \inferrule{\null}{\emptycontext\vdash}{\regle{ec}}  & \inferrule{\GG\vdash
                                                          A}{\GG,x:A\vdash}{\regle{ce}}
                                                          \quad \text{Where
                                                          $x\notin\Var\GG$}\\
    \multicolumn{2}{l}{\text{\emph{For types}:}}\\
    \inferrule{\GG\vdash}{\GG\vdash\Obj}{\regle{$\Obj$-intro}} & \inferrule{\GG\vdash t:A~\\ \GG\vdash u:A}{\GG\vdash\Hom Atu}{\regle{$\Hom{}{}{}$-intro}}\\
    \multicolumn{2}{l}{\text{\emph{For terms}:}}\\
    \inferrule{\Gamma\vdash \\ (x : A) \in \GG}{\Gamma\vdash x:A}{\regle{var}} \\
    \multicolumn{2}{l}{\text{\emph{For substitutions}:}}\\
    \inferrule{\GD\vdash}{\GD\vdash \sub{}:\emptycontext}{\regle{es}} & \inferrule{\GD\vdash \Gg: \GG \\ \GG,x:A\vdash~\\ \GD\vdash t:A[\Gg]}{\GD\vdash\sub{\Gg,x\mapsto t}:(\GG,x:A)}{\regle{se}} \\
    \multicolumn{2}{l}{\text{\emph{Definitional equality}:}}\\
    \inferrule{\GG\vdash t:A \\ \GG\vdash A\equiv B}{\GG\vdash t:B} &
                                                                          \inferrule{\GG\vdash
                                                                          t:\unit
                                                                          \\
        \GG\vdash u:\unit}{\GG\vdash t\equiv u:\unit}{(\eta_{\unit})}
  \end{array}
\]

\paragraph{Semantics.}
\begin{align*}
  \Syn{\Glob_{\unit}} &= \op{\FinGSet}\\
  \Mod{\Syn{\Glob_{\unit}}} &= \GSet
\end{align*}

\subsection{The theory $\MCaTT$}\label{app:mcatt}
\paragraph{Syntax.}
\begin{induction}
\item \emph{Contexts}: lists $(x_0:A_0,\ldots,x_n:A_n)$ with $x_i$ variables and
  $A_i$ types
\item \emph{Types}: either $\unit$ or of the form $\Hom Atu$ with $A$ type and
  $t$ and $u$ terms
\item \emph{Terms}: either variables or of the form $\mop_{\GG,A}[\Gg]$ or
  $\mcoh_{\GG,A}[\Gg]$ with $\GG$ a ps-context, $A$ a type and $\Gg$ a
  substitution
\item \emph{Substitutions}: lists $\sub{x_0\mapsto t_0,\ldots,x_n\mapsto t_n}$
  with $x_i$ variables and $t_i$ terms
\end{induction}

\paragraph{Desuspension.}
\[
  \begin{array}{cc}
    \text{\emph{For the context $\emptycontext\vdash$}} &
                                                          \text{\emph{For the context $(\GG,x:A)\vdash$}} \\
    \chop\emptycontext = \emptycontext & \chop{(\GG,x:A)} = \left\{
                                         \begin{array}{ll}
                                           \chop{\GG} & \text{if $A~= \Obj$} \\
                                           \chop{\GG} , x:\chop{A} & \text{otherwise}
                                         \end{array}
                                                                     \right.
    \\[1.2em]
    \text{\emph{For the type $\GG\vdash\Obj$}} &
                                                 \text{\emph{For the type
                                                 $\GG\vdash\Hom Atu$}} \\
    \chop{\Obj} = \Obj & \chop{(\Hom Atu)} = \left\{
                         \begin{array}{ll}
                           \Obj & \text{if $A~= \Obj$} \\
                           \Hom{\chop A}{\chop t}{\chop u} & \text{otherwise}
                         \end{array}
                                                             \right.
    \\[2.2em]
    \text{\emph{For a variable $\GG\vdash x:A$}} &
                                                   \text{\emph{For the term
                                                   $\GD\vdash \cohop_{\GG,A}[\Gg] :
                                                   A[\Gg]$}} \\
    \chop x = x &  \chop{\cohop_{\GG,A}[\Gg]} = \mop_{\GG,A}[\chop \Gg] \\[1.2em]
                                                        &
                                                          \text{\emph{For the term $\GD\vdash \coh_{\GG,A}[\Gg] : A[\Gg]$}} \\
                                                        &
                                                          \chop{\coh_{\GG,A}[\Gg]} =
                                                          \mcoh_{\GG,A}[\chop\Gg]\\[1.2em]
    \text{\emph{For $\GD\vdash\sub{}:\emptycontext$}} &
                                                        \text{\emph{For
                                                        $\GD\vdash\sub{\Gg,x\mapsto
                                                        t}:(\GG,x:A)$}} \\
    \chop{\sub{}} = \sub{} & \chop{\sub{\Gg,x\mapsto t}} = \left\{
                             \begin{array}{ll}
                               \chop \Gg & \text{if $A~= \Obj$} \\
                               \sub{\chop \Gg, x\mapsto \chop t} & \text{otherwise}
                             \end{array}
                                                                   \right.
  \end{array}
\]

\paragraph{Inference rules.}
\[
  \begin{array}{cc}
    \multicolumn{2}{l}{\text{\emph{For contexts} :}}\\
    \inferrule{\null}{\emptycontext}{\regle{ec}}
    & \inferrule{\GG\vdash A}{\GG,x:A\vdash}{\regle{ce}}\\
    \multicolumn{2}{l}{\text{\emph{For types} :}}\\
    \inferrule{\GG\vdash}{\GG\vdash \unit}{\regle{$\unit$-intro}}
    & \inferrule{\GG\vdash A \\ \GG\vdash t:A \\ \GG\vdash u :A}{\GG\vdash\Hom Atu}{\regle{$\Hom{}{}{}$-intro}}\\
    \multicolumn{2}{l}{\text{\emph{For terms} :}}\\
    \inferrule{\GG\vdash \\ (x:A)\in\GG}{\GG\vdash x:A}{\regle{Var}}
    &\inferrule{\GG\vdashps \\ \GG\vdashop A \\ \GD\vdash \Gg : \chop\GG}{\GD\vdash\mop_{\GG,A}[\Gg]:\chop A[\Gg]}{\regle{$\mop$-intro}}\\
    \inferrule{\GG\vdash}{\GG\vdash ():\unit}{\regle{$()$-intro}}
    & \inferrule{\GG\vdashps \\ \GG\vdasheq A \\
    \GD\vdash\Gg:\chop\GG}{\GD\vdash\mcoh_{\GG,A}[\Gg]:\chop
    A[\Gg]}{\regle{$\mcoh$-intro}}\\
    \multicolumn{2}{l}{\text{\emph{For substitutions} :}}\\
    \inferrule{\GD\vdash}{\GD\vdash\sub{}:\emptycontext}{\regle{es}} &
                                                                       \inferrule{\GD\vdash
                                                                       \Gg:\GG
    \\ \GG\vdash A \\ \GD\vdash t:A[\Gg]}{\GD\vdash\sub{\Gg,x\mapsto
    t}:(\GG,x:A)}{\regle{se}} \\
    \multicolumn{2}{l}{\text{\emph{Definitional equality} :}} \\
    \inferrule{\GG\vdash t:A \\ \GG\vdash A\equiv B}{\GG\vdash t:B} &
                                                                      \inferrule{\GG\vdash
                                                                      t:\unit
    \\
    \GG\vdash u:\unit}{\GG\vdash t\equiv u:\unit}{(\eta_{\unit})}
  \end{array}
\]

\paragraph{Semantics.}
\begin{itemize}
\item $\Syn{\MCaTT}$ is equivalent to the full subcategory of $\Syn{\CaTT}$
  whose objects are the contexts with a unique variable of type $\Obj$.
\item $\Mod{\Syn{\MCaTT}}$ is equivalent to the full subcategory of
  $\Mod{\Syn{\CaTT}}$ that define a single $0$-cell
\end{itemize}

\end{document}